\newcommand\ha{{H$\alpha$}}
\newcommand\hb{{H$\beta$}}
\newcommand\PaB{{Pa$\beta$}}

\newcommand\LUM{\:{\rm erg\:s^{-1}}}
\newcommand\FLUX{\:{\rm erg\:cm^{-2}\:s^{-1}}}
\newcommand\FLUXARCSEC{\:{\rm erg\:cm^{-2}\:s^{-1}\:arcsec^{-2}}}

\newcommand\VEL{\:{\rm km\:s^{-1}}}

\newcommand\OiL{[\ion{O}{1}] $\lambda 6300$}

\newcommand\NiiL{[\ion{N}{2}] $\lambda\lambda 6548, 6583$}

\newcommand\FeiiL{[\ion{Fe}{2}] 1.644 $\mu$m}
\newcommand\FeiiLL{[\ion{Fe}{2}]\,1.257 $\mu$m}
\newcommand\PaBL{Pa$\beta$\,1.28 $\mu$m}

\newcommand\sii{[\ion{S}{2}]}
\newcommand\nii{[\ion{N}{2}]}

\newcommand\feii{[\ion{Fe}{2}]}
\newcommand\hii{\ion{H}{2}}

\newcommand\gal{NGC\,6946}

\newcommand{\EXPU}[3]{\mbox{\rm $#1 \times 10^{#2} \rm\:#3$}}  

\documentclass[modern]{aastex63}

\shorttitle{SNRs in NGC~6946}
\shortauthors{Long, Blair, Winkler, \& Lacey}

\usepackage{natbib}
\usepackage{comment}
\usepackage{todonotes}
\usepackage{graphicx}
\begin{document}


    \title{The Supernova Remnant Population of NGC\,6946 as Observed in [Fe~II]\ 1.644 $\mu$m with HST\footnote{Based on observations with the NASA/ESA Hubble Space Telescope obtained at the Space Telescope Science Institute, which is operated by the Association of Universities for Research in Astronomy, Incorporated, under NASA contract NAS5-26555. Support for program numbers 14638 and 15216 was provided through a grant from the STScI under NASA contract NAS5-26555}}

\correspondingauthor{Knox S. Long}
\email{long@stsci.edu}

\author[0000-0002-4134-864X]{Knox S. Long}
\affil{Space Telescope Science Institute,
3700 San Martin Drive,
Baltimore MD 21218, USA; long@stsci.edu}
\affil{Eureka Scientific, Inc.
2452 Delmer Street, Suite 100,
Oakland, CA 94602-3017}

\author[0000-0003-2379-6518]{William P. Blair}
\affiliation{The Henry A. Rowland Department of Physics and Astronomy, 
Johns Hopkins University, 3400 N. Charles Street, Baltimore, MD, 21218; 
wblair@jhu.edu}

\author[0000-0001-6311-277X]{P. Frank Winkler}
\affiliation{Department of Physics, Middlebury College, Middlebury, VT, 05753; 
winkler@middlebury.edu}

\author[0000-0001-9192-517X]{Christina K. Lacey}
\affiliation{Hofstra University, NY, USA; christina.lacey@hofstra.edu}

\begin{abstract}
 
NGC\, 6946 is a high star formation rate face-on spiral galaxy that has hosted ten supernovae since 1917.  Not surprisingly, a large number of supernova remnants and candidates have been identified either as optical nebulae with high \sii:\ha\ line ratios (147) or as compact non-thermal radio sources (35).
However, there are only seven overlaps between these two samples. Here, we apply \FeiiL\ emission as a new diagnostic to search for supernova remnants in an attempt to resolve this discrepancy. 
\feii\ is expected to be relatively strong in the radiative shocks of supernova remnants and almost absent in \hii\ regions. It is less susceptible to the effects of  absorption along the line of sight than the optical lines normally used to identify remnants. Using data from the WFC3 camera on {\em HST}\@,
we identify 132 \feii\ emission nebulae in NGC\,6946 as likely supernova remnants. Of these, 54 align with previously known optical supernova remnants.  The remaining 78 objects are new; of these 44 are visible in new {\em HST} imagery in \ha\ and \sii.  This brings the total number of supernova remnant candidates (from optical and/or IR data) in NGC\,6946 to 225.
A total of 14 coincidences with radio supernova remnant candidates (out of 30 in our search area) are found in this expanded list.  The identification of so many new remnant candidates validates the use of \feii\ imagery for finding remnants, and suggests that previous remnant searches in other galaxies may be far from complete.  


\end{abstract}

\keywords{galaxies: individual (NGC~6946) -- galaxies: ISM  -- supernova remnants}

\section{Introduction} \label{sec:intro}

Supernova remnants (SNRs) distribute the ashes of exploded stars into the interstellar medium, thus recycling their processed material into the next generation of stars and planets.  SNRs are inherently multiwavelength emitters, and their appearance in various wave bands reflects both the nature of the progenitor and the environment into which they are expanding.  SNRs in external galaxies are thought to be the best way to understand SNRs as a class, because all of the objects in a galaxy are effectively at the same distance, and the appearance of the objects is less affected by absorption within a galaxy than is the case in the Milky Way, especially for galaxies that are viewed approximately face-on. Also, the characteristics of the population of SNRs within a given galaxy is related to the star formation rate and other processes like the overall ISM pressure that affect the evolution of the galaxy itself.

For reasons associated with sensitivity and angular resolution, most SNRs in external galaxies have been first identified through narrow-band optical imaging, \cite[see, e.g.][for a review]{long17}.  Optically, SNRs are emission nebulae with \sii:\ha\ ratios $\gtrsim 0.4$ compared to \hii\ regions which, at least for high surface brightness nebulae, have \sii:\ha\ ratios $\lesssim 0.2$\@.  Smaller numbers of extragalactic SNRs have been identified as non-thermal radio sources \citep[associated with \ha\ emission to avoid contamination by background sources; see][]{lacey01}.  While in the Magellanic Clouds, and recently M33, most SNRs have now been  detected both optically and at radio wavelengths, this is not generally the case in more distant galaxies---although this is beginning to change as radio sensitivities improve. Usually radio-identified SNRs are in the brightest portions of the spiral arms of a galaxy, where optical identification of SNRs can be relatively difficult because of overlying dust and emission from \hii\ regions;  this is a plausible explanation for the fact that many radio SNR candidates have so far not been detected optically.

The nearby \citep[7.8 Mpc,][]{murphy18, anand18},\footnote{In \citet{long19}, we used a distance of 6.72 Mpc, based on the work of \cite{tikhonov14}, but more recent tip-of-the-red-giant-branch measurements    by \citet[][$7.83 \pm 0.29$ Mpc]{murphy18}   and  \citet[][$7.72\pm 0.32$ Mpc]{anand18}, both based on {\em HST} data,   suggest that 7.8 Mpc is more accurate.  All the values in this paper for size, luminosity, etc., of objects in \gal\ assume this distance.} nearly face-on \cite[$i = 32.6\degr$,][]{deblok08} galaxy \gal\ is a case in point.  With a total star formation rate of at least  $3.2\, M_\sun\,{\rm yr}^{-1}$  \citep{jarrett13} and perhaps as high as 12.1 $M_\sun\,{\rm yr}^{-1}$ \citep{eldridge19}, it has hosted 10 supernovae (SNe) since 1917, the most recorded in any galaxy.\footnote{This excludes the so-called failed SN discovered by \cite{adams17}.}   Thus one might expect \gal\ to have of order 1000 SNRs with ages less than 10,000 yr (if indeed they stay visible for this long).   The first 27 of these SNRs were identified by \citet[][hereafter MF97]{matonick97} using the optical \sii:\ha\ ratio criterion.  Shortly thereafter, \cite{lacey01} identified 35 SNR candidates in the galaxy as non-thermal radio sources, only two of which was in the MF97 list.\footnote{One of these two, MF16 = L97-85, has subsequently been shown to be an ultraluminous X-ray source; see discussion in Sec.~4.4.}   More recently, we conducted a new (ground-based) optical search for SNRs in \gal\ \cite[][hereafter L19]{long19}, where we identified 147 SNR candidates (including the 27 from MF97) based on high \sii:\ha\ ratios in narrow-band imaging.  As part of the same study we obtained spectra of 102 of the candidates in an attempt to confirm the flux ratios (and thus the SNR identifications), and confirmed that 89 candidates (87\% of the those observed)  did indeed have \sii:\ha\ ratios greater than 0.4.  Of the 147 candidates, seven were coincident with radio SNR candidates from \cite{lacey01}; six of these seven have observed spectra in L19, and five of these six  were confirmed to have high \sii:\ha.  Having at least five of the radio SNR candidates with confirmed optical SNRs in our deeper optical survey suggests that even more sensitive (or different types of) observations might confirm more of the radio objects as SNRs.

An alternative, but largely unexplored, way to identify SNRs in nearby galaxies is to search for emission nebulae in the light of \FeiiL.
Observations of Galactic and LMC SNRs show  \FeiiL\,:\,\PaB~1.25 $\mu$m 
ratios ranging from about 0.7 in young SNRs to $\sim 10$ in older objects \citep{mouri00,labrie06,koo15}. By contrast, this ratio is generally $<0.1$ in \hii\ regions and other photoionized nebulae \citep{lowe79,armand96,reiter19}.   An important advantage of \FeiiL\  is that it is less sensitive to foreground absorption, which is important in the case of \gal\ since it lies only about 11\fdg 7 out of the Galactic plane, and has a foreground extinction of 
$A_{{\rm V}}  =  0.95$ mag \citep{schlegel98,schlafly11}\footnote{As reported in the NASA/IPAC Extragalactic Database (NED), which is funded by the National Aeronautics and Space Administration and operated by the California Institute of Technology.}. Also, as with many face-on spirals, substantial regions of extended dust absorption are intrinsic to the galaxy itself, which could affect the detectability of SNRs with the normal optical criterion.

The near-IR \FeiiL\ line was first detected in the Galactic SNR MSH11-5{\em 2} by \citet{seward83}.  Spectra obtained by \cite{graham87} and \cite{oliva89} of other Galactic and Magellanic Cloud SNRs showed that \FeiiL\ was bright compared to the near-IR lines of hydrogen, such as Br$\gamma$ and Pa$\beta$.  A  number of imaging and spectroscopic observations of \FeiiLL\ and \FeiiL\ for SNRs in nearby galaxies, including M33 \citep{lumsden95,morel02}, NGC\,253 \citep{forbes93}, and M82 \citep{greenhouse91,greenhouse97} followed, with much of the interest focused on trying to use the integrated diffuse \feii\ emission to estimate the SN rate in more distant (often obscured) galaxies \cite[see, e.g.][]{rosenberg12}. There was very little follow-up of these early studies of SNRs, however, mostly due to limitations in instrumental sensitivity and telluric contamination for near-IR ground-based work.  The development of modern detectors and improved techniques is beginning to change this.  Recently, for example, \cite{lee19} have detected about 25\% of the known SNRs in the first quadrant of the Galaxy in \FeiiL\ narrow-band images, and \citet{blair14} detected a large number of M83 SNRs  in \FeiiL\ using narrow-band imaging with the IR camera on {\em HST}/WFC3. 

Recently, \cite{bruursema14}  made an attempt to identify SNRs in NGC\,6946  in the light of  \FeiiL\  using the WIYN High Resolution Infrared Camera on the WIYN 3.5 m telescope.  They identified 48 candidate \feii\ objects, very few of which were associated with previous optical or radio SNR candidates.  Here we describe a new, much more sensitive attempt to identify SNRs using narrow-band imaging of \FeiiL\ with the WFC3 IR camera on {\em HST}\@.  Our primary motivations for this current study were to provide a more complete sample of SNRs in NGC\,6946 and to see whether we could confirm a larger number of the radio-identified SNR candidates as actual SNRs.   To further advance our knowledge of the SNR population in NGC\,6946, we have also recently obtained WFC3 UVIS \ha\ and \sii\ imaging data as well, for which we give a preliminary report; these optical data enable us to measure far more precise sizes for the SNRs than had been possible from previous ground-based surveys by ourselves and others, as well as allowing us to identify some new optical candidates that align with compact \feii\ sources.

The current paper is organized as follows:  In Sec.\  \ref{sec_observations}, we describe both the IR and UVIS observations and our reduction methods. 
In Sec.\ \ref{sec_catalog}, we report the development of our catalog of \feii\ SNR candidates, and in Sec.\ \ref{sec_analysis} we provide a comparison of these objects to the existing catalogs of SNRs.  In Sec.\ \ref{sec_discussion}, we discuss the implications of these comparisons.  Finally, in Sec.\ \ref{sec_summary}, we summarize the results and suggest paths forward for using NIR \feii\ imaging and spectroscopy to better understand SNRs as a class.

\section{Observations and Data Reduction \label{sec_observations}}

\begin{figure}
\plotone{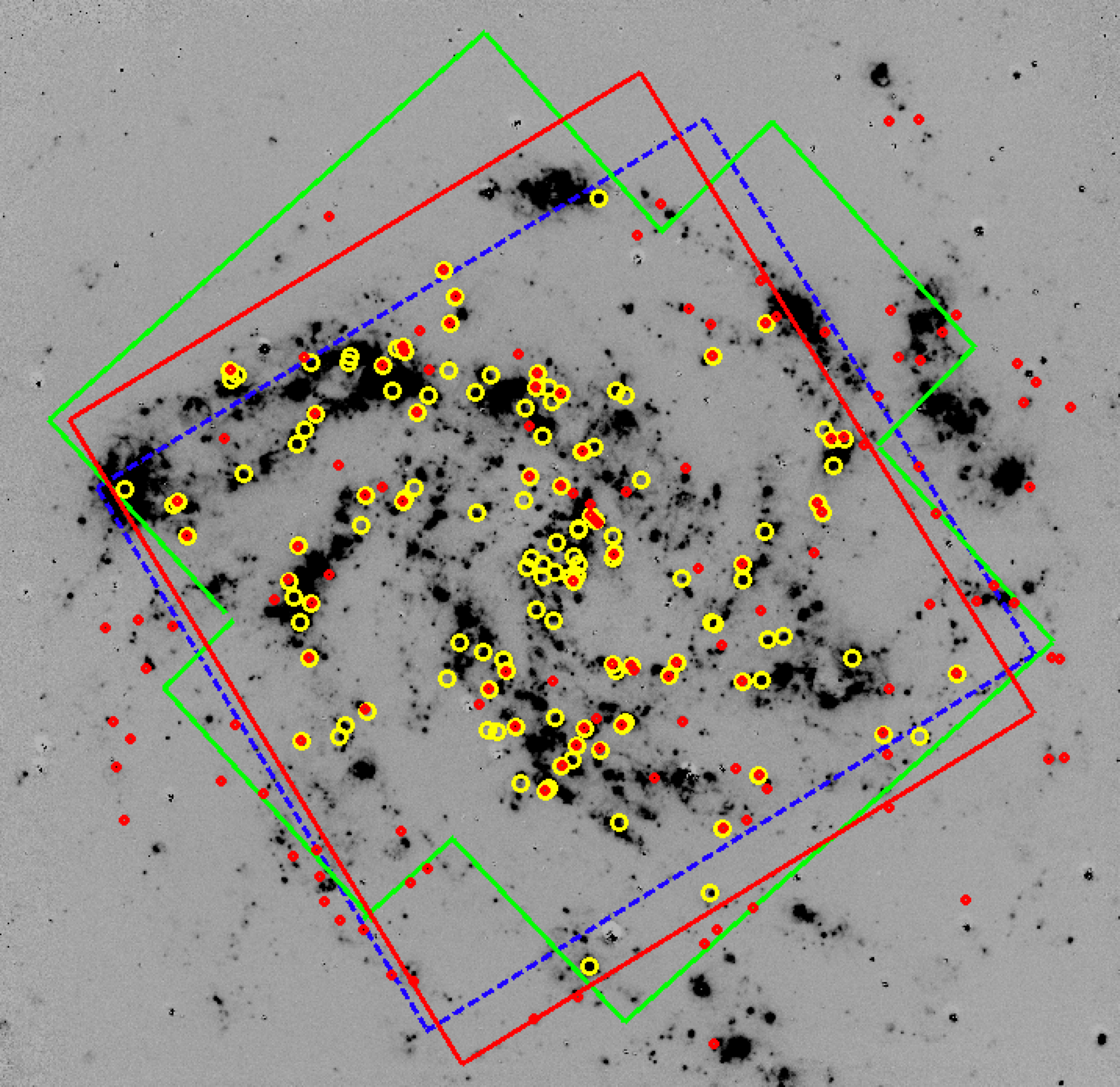}
\caption{A continuum-subtracted \ha\ image of \gal\  from the WIYN 3.5 m telescope on Kitt Peak (see \citep{long19}.  The red rectangle shows the 9-field WFC3 IR region covered with the F164N and F160W filters; the green region shows the footprint of the seven UVIS fields observed in \ha, \sii, and F547M; and the dashed blue rectangle shows the region covered in archival  F128N (Pa$\beta$) and F110W from {\em HST} program 14156 (PI Leroy). Red circles show the optical SNR candidates identified by \citet{long19},  and yellow circles show the  \feii\ sources 
found in the independent (blind) search of the WFC3 IR data. (Many of the sources  overlap; see text.) 
\label{fig_overview}}
\end{figure}

The primary observations for this project (Program ID 14638, Long - PI) took place on 2016 October 26-28  and were carried out with the IR camera on {\em HST}'s WFC3.  The observations comprised a $3\times 3$ mosaic of pointings, covering most of the inner parts of the galaxy as shown in Fig.\ \ref{fig_overview}.  Each field of the mosaic was imaged in \feii\ with the  F164N filter for 2400s  and in an overlapping continuum band with the F160W filter for 600s---both in a single orbit.  The four exposures associated with each filter in each field were dithered to minimize the effects of bad pixels and to improve the characteristics of the images for constructing images with AstroDrizzle.   The mosaic was oriented to cover much the same region of NGC\,6946 that had been imaged earlier with WFC3 in the light of \PaB\ (F128N) and in the broad band F110W filter (Program ID 14156, Leroy PI).  In both cases, the continuum-band observations were intended primarily to permit the creation of continuum-free emission-line images, but they are also useful for investigating the local stellar populations, a topic we defer to a future paper.

In order to carry out the project, we have also made use of observations of NGC\,6946 (Program ID 15216, Blair - PI)  made with the UVIS camera on WFC3, using the F657N (\ha+\nii),  F673N (\sii) and F547M continuum filters. Comprising seven overlapping fields, the coverage for these data is shown in green in Fig.\ \ref{fig_overview}.
The WFC3 UVIS data were obtained on 2019 January 25-27.  Each of the seven fields was observed for 2826 s in F657N, 3853 s in F673N, and 1473 s in F547M.  Each of the UVIS data sets was dithered in three steps to remove cosmic rays and cover the chip gap, A FLASH parameter of 10, 10, and 7, respectively, was used to reduce effects of  charge transfer inefficiency.  As was the case for the IR continuum-band observations, the F547M continuum band observations were intended primarily to permit the creation of continuum-free emission-line images by subtracting out the galaxy and stellar background.  

The purposes of the UVIS observations were  (a) to better characterize the known SNR population through precise size measurements,
(b) to identify SNRs missed in the ground-based search of for SNRs in NGC\,6946---especially useful for small-diameter remnants and ones in confused regions, and (c) to further investigate the stellar populations in the vicinity of the SNRs.  In this paper we use the UVIS data primarily to aid in characterizing the \feii\ SNR candidates, and to compare them with the optical SNRs.   A more complete analysis of the WFC3 UVIS study will be published separately.

We retrieved all of the above data sets in the summer of 2019 to obtain the latest calibrations\footnote{These data are available from MAST at \url{https://www.doi.org/10.17909/t9-9skz-qw10}} .  We then created mosaicked images in each filter band using the AstroDrizzle package \citep{fruchter10}.  Our approach was first to update (slightly) image world coordinate system (WCS) information by aligning the pipeline-produced drizzled images of each field with Gaia DR2 stars \citep{gaia-collaboration18}.  We then applied the new WCS from the drizzled images to each individual exposure for the set of images taken with each filter.  For the IR observations, we then created a mosaicked image for the F160W filter and used that as a reference image with AstroDrizzle to create mosaicked images for the F110W, F128N and F164N filters. We followed a similar procedure for the UVIS images, using the mosaicked F547M image as the reference image in that case. The resulting mosaic images are all aligned onto the same grid and pixel scale, allowing direct comparison of IR and UVIS images. {There was very little difference among the sizes of the point spread functions for the resultant line and continuum images in either the UVIS or IR images.  There was, of course, a difference between the UVIS and IR images, reflecting both the diffraction limit and the pixel scales (0\farcs 13 for the IR camera, 0\farcs 04 for the UVIS) in the two wavelength regimes.}

The alignment from this procedure allows us to carry out simple arithmetic on the various images to produce ``pure'' emission-line images.   The images through both of the IR continuum filters (F110W and F160W) are broad enough that they include the emission lines of interest (\PaBL\ and \FeiiL, respectively), so it was necessary to correct for this fact.  To do so, consider images through a pair of line and continuum filters, e.g., $L =  $ F164N and $C = $ F160W:

\begin{equation}
C_{obs}=C+aL\, ,
\end{equation}
where $a$ is the fractional contribution of the emission-line filter to the observed continuum image, $C_{obs}$. Similarly, what is observed in the narrow-band image $L_{obs}$ is

\begin{equation}
L_{obs}=bC+L\,,
\end{equation}
where $b$ is the fractional contribution of the true continuum image to the observed image.
We can can solve for the ``true'' values of the line and continuum, viz.

\begin{equation}
C=\frac{C_{obs}-a\:L_{obs}}{1-ab}\,,
\end{equation}

and

\begin{equation}
L=\frac{L_{obs}-b\:C_{obs}}{1-ab}\, .
\end{equation}
Thus, the ``pure'' continuum image can be used to subtract the emission-line images, producing accurate subtractions that have minimal impact on the emission line fluxes.  We used these ``pure'' continuum and ``pure'' emission-line images in what follows.

For the UVIS filters, \sii\ and \ha, the F547M filter is free of any significant emission-line contamination, so one can simply derive the proper scaling factors and directly subtract the F547M image from the F673N and F657N filters to obtain pure \sii\ and \ha\ images, respectively.\footnote{Technically, the F657N filter passes some emission from \NiiL\ as well as \ha, but \ha\ dominates in these data.  For brevity, we refer to these data as the ``\ha''  data. } 

For both IR and UVIS data, the color variations of the stellar background and the relatively broad continuum filters combine to make a perfect continuum subtraction impossible. Practically speaking, however, the continuum subtraction is more than sufficient to greatly enhance the line emission and make valid comparisons possible.  For regions that over-subtract, the negative residuals are obvious in the data and can be ignored; stars that under-subtract leave residuals that could be mistaken for very small diameter emission sources, but these can be identified by careful comparison to the unsubtracted continuum image during visual inspection.  For inspecting the images visually, we tend to slightly over-subtract the continuum to minimize the chance of mistaking a stellar residual as a real emission object.

\section{A catalog of [Fe~II] emission nebulae \label{sec_catalog}}

To obtain a list of \feii\ sources, we carried out multiple visual searches, comparing apparent compact emission excesses in the subtracted \feii\ data to other images to confirm their voracity.  First, using the images of individual fields and a preliminary version of the multi-field mosaic, three of us (KSL, WPB, and PFW) carried out  independent blind searches, comparing the subtracted \feii\ images and the corrected continuum images (in order not to be misled by stellar residuals).  Comparing our independent lists, we found that the majority of objects appeared in all three lists; for the remaining objects we arrived at a consensus decision to keep or reject them.  

Then, following reprocessing with AstroDrizzle, which resulted in superior mosaic images and improved subtractions, we repeated the process:  Two of us (WPB and PFW) carried out entirely new blind searches, where we compared the continuum-subtracted \feii\ mosaic not only with the corrected F160W continuum, but also with the continuum-subtracted WFC3 \ha\ and \sii\ mosaics. Once again, the agreement between our two lists was high, and we arrived at consensus on the differences through a joint visual comparison.  Finally, we compared our new list to the preliminary blind-search list and found 11 objects that did not appear in the new list.  Visual inspection of those positions resulted in our accepting four of these 11 as additional valid candidates.   

We individually graded the candidates as  A = solid \feii\ candidates; B = more marginal candidates due to some continuum confusion and/or faintness; or C = ``objects" showing some \feii\ emission that we regarded as unlikely to be true \feii\ detections (either likely stellar residuals or objects only marginally above the background).   Our final catalog, shown in Table \ref{ngc6946_tab1}, contains 132 objects -- 93  of which were graded as ``A" and 39 as ``B"; we do not report ``C" graded objects.  Of the 147 optical SNRs reported by L19, 92 are within the footprint of the WFC3 \feii\ data, and 54 of these (59\%) had counterparts in our \feii\ blind search.

After the blind search and verifications were done, we were then able to search the \feii\ source positions for optical counterparts that were visible in the {\em HST} UVIS data but that had not been found from the ground-based search (L19).  From this exercise, we found the following: (a) 22 sources with compact and/or faint optical counterparts that had been missed in the ground-based searches; (b) 22 sources where the \feii\ source was projected amidst \hii\ emission, but for which the {\em HST} images enabled us to identify a likely optical candidate; (c) 13 cases where a well-defined \feii\ source was projected against \hii\ emission but no specific optical counterpart could  be identified; and (d) 21 sources with no optical counterpart, but which are located in obvious dust lanes or patches.

In the following section, we present some examples of the categories discussed above.

\subsection{Examples}

\begin{figure}
\plotone{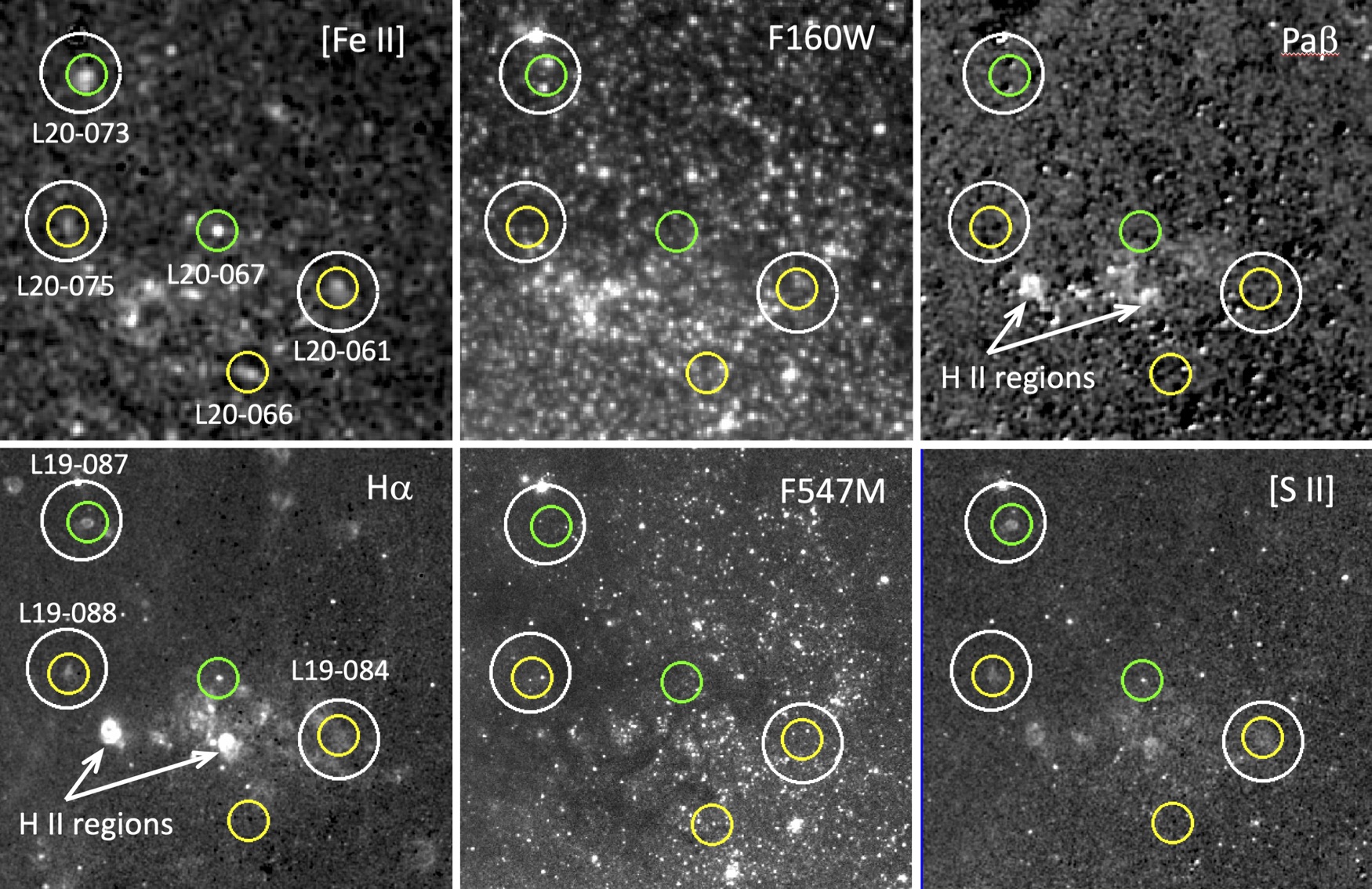}
\caption{In this and the following two figures, we show 6-panels with examples of the identified \feii\ sources in various {\em HST}/WFC3 wavebands.  The six panels include (from upper left, top row): subtracted IR \feii, IR F160W, subtracted IR \PaB, and bottom row: subtracted UVIS \ha, UVIS F547M, and subtracted UVIS \sii.  Small circles are 2\arcsec\ in diameter and indicate \feii\ sources identified in the blind search, as identified in the \feii\ panel, with green and yellow denoting `grade A' and  `grade B,' respectively (see text).  The large circles are 4\arcsec\ in diameter and indicate the positions of optical SNR candidates as cataloged in \cite{long19}. Large white circles are optical SNRs with \feii\ counterparts in the blind search, and red circles (following figures) are optical SNRs that did not have \feii\ counterparts from the blind search. The data are shown with log scaling to increase the dynamic range, and some frames have been smoothed slightly for display, to diminish the pixelated noise in the background.  Note that none of these SNRs shows detectable \PaB\ emission, although nearby \hii\ regions (indicated) do. See text for further explanation.
\label{fig_example1}}
\end{figure}

\begin{figure}
\plotone{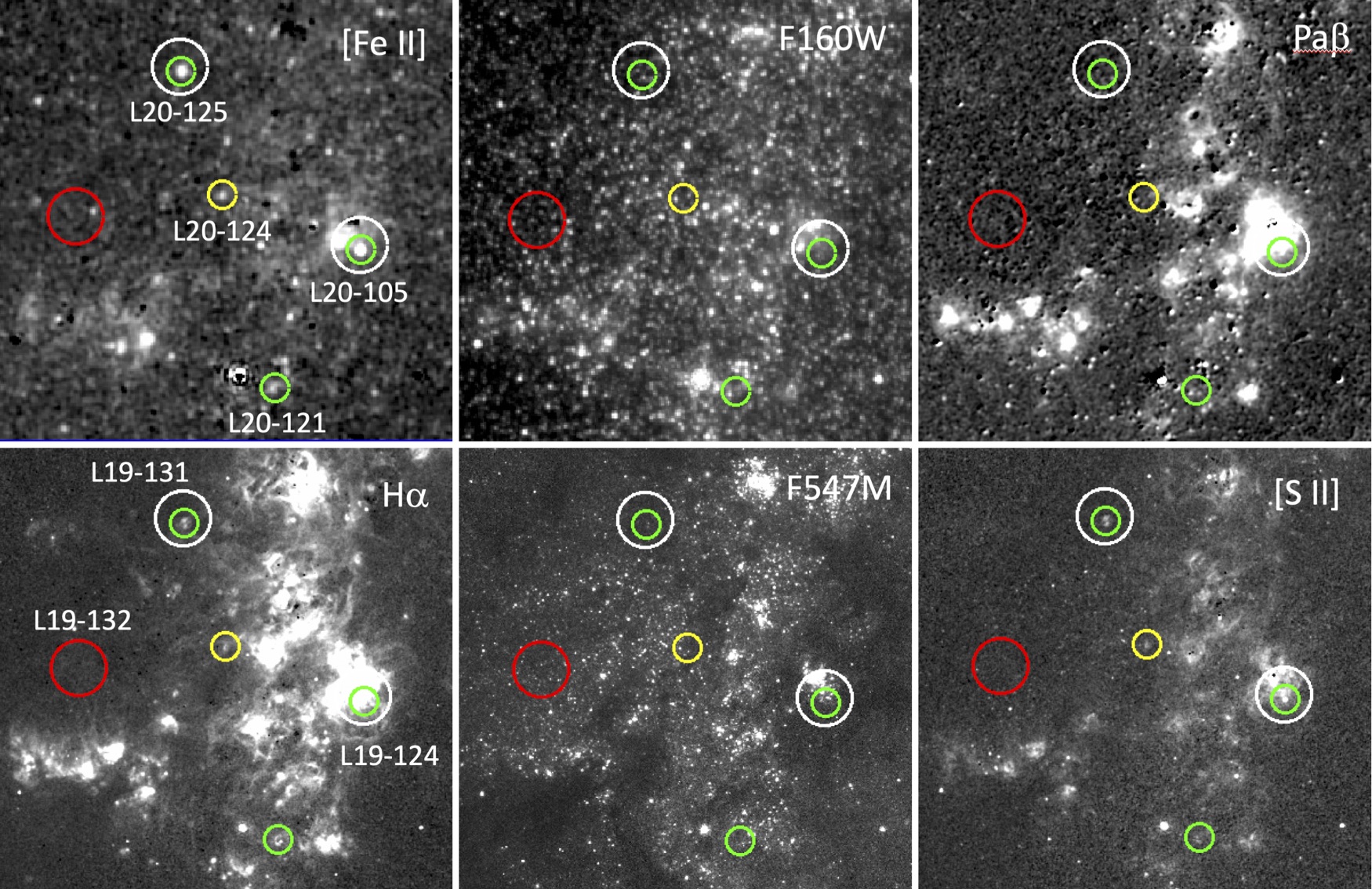}
\caption{Same as Fig.\ \ref{fig_example1} but for a complex  region along the eastern spiral arm that has many bright \hii\ regions. The red circle shows the position of an optical SNR identified in the L19 survey, but whose surface brightness is too low be be seen with WFC3 (either IR or UVIS) at the current level of exposure.  Note how well SNR L19-124 (=\,L20-105) stands our on the \feii\ panel despite its proximity to the bright \hii\ region to the NE\@.  See text for further explanation of this field.
\label{fig_example2}}
\end{figure}

\begin{figure}
\plotone{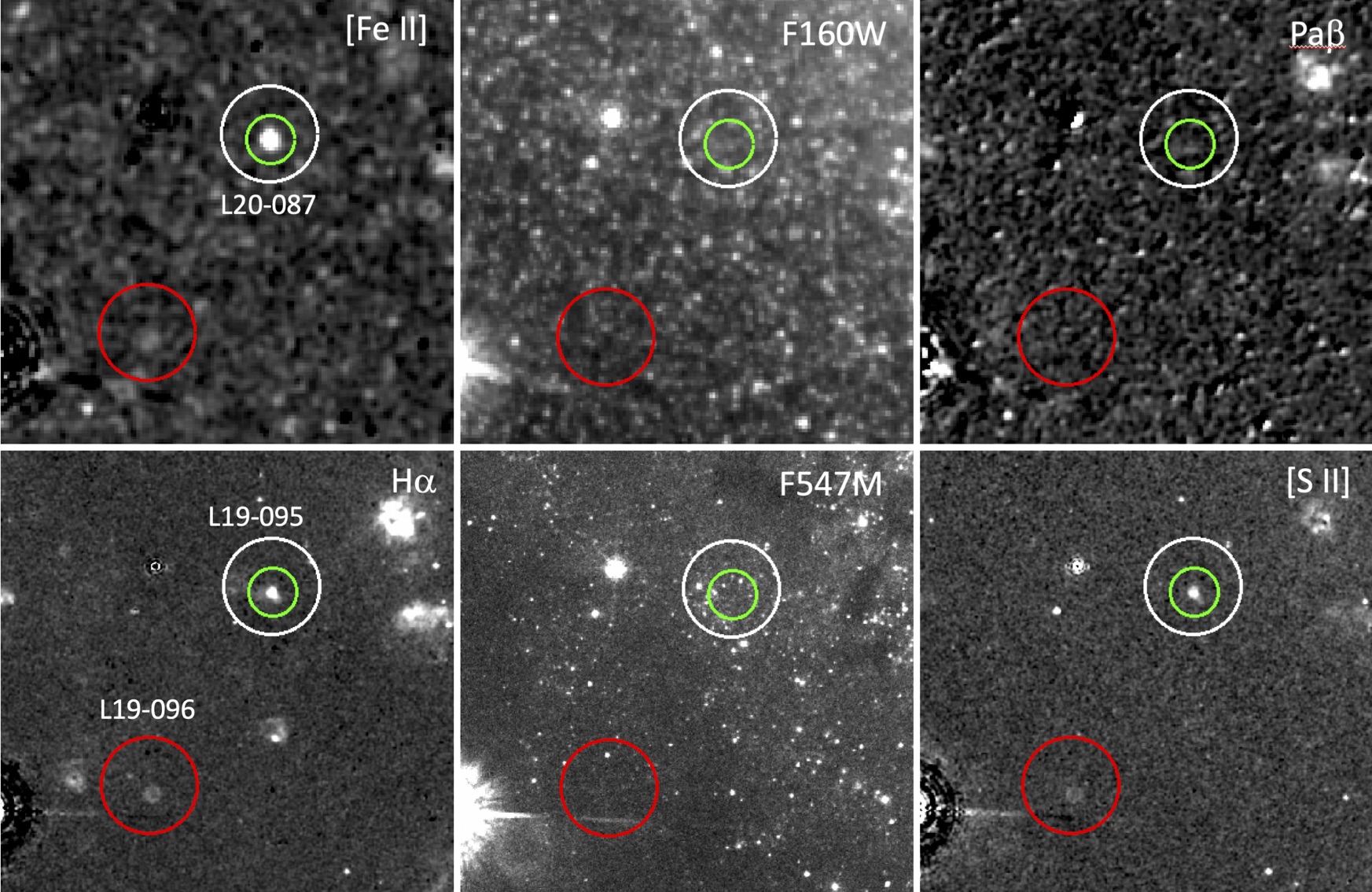}
\caption{Same as Fig.\ \ref{fig_example1} but for two optical SNR candidates, L19-095 and L19-096\@. L19-095 (=\,L20-087) is a compact, bright SNR in the optical emission lines, and is well-detected but looks larger in the \feii\ frame due to the larger pixels in the IR camera.  The fainter L19-096 is apparently also detected in \feii\ but was missed in our blind search.  Note the bright, compact \hii\ region at upper right that shows \PaB, while the SNRs do not.  Possible faint \feii\ may be present from the \hii\ region, but at a level much lower than \PaB\ or \ha\ emission.
\label{fig_example3}}
\end{figure}

The \feii\ nebulae we identified are fairly diverse, as shown in  Figs.\  \ref{fig_example1} - \ref{fig_example3}.
Fig.\ \ref{fig_example1} shows a small region of \gal\ in multiple {\em HST} WFC3 bands with a sampling of representative objects. Small circles indicate \feii\ objects identified in our blind search:  green circles for  A-graded objects and yellow circles for B-graded ones.  Larger circles show optical SNR positions from L19.  Three of the five \feii\ objects align with known optical SNRs.  The `A' source L20-067 aligns with a very compact optical SNR seen in the WFC3 \ha\ and \sii\ frames but that was {\em not} identified in the L19 ground-based search; presumably the light from this compact source was smeared out and confused with the \hii\ complex to the south and east at ground-based resolution, and it was thus not identified. It stands out very well in the \feii\ panel.  The `B' source L20-066 does not show an optical SNR counterpart,  due to its faintness, or possibly because it suffers from high extinction.  None of the objects has severe stellar or \hii\ contamination, and none of the SNRs has detectable Pa$\beta$ emission, although two nearby compact \hii\ regions (indicated) are clearly present in \PaB. This is a general pattern seen in the data where, at the current survey depth, almost no \PaB\ counterparts are clearly seen for the compact \feii\ sources, while \PaB\ is visible in the nearby \hii\ regions.

In Fig.\ \ref{fig_example2} we show a more complicated region of the eastern spiral arm of \gal.  Despite the numerous stellar residuals in the subtracted \feii\ image, with careful comparison against the continuum images we were able to identify four \feii\ sources. Two of these objects align with L19 SNRs. 
For one of these, L19-124, nearby \hii\ contamination made it   difficult to identify in our ground-based optical survey, but it  stands out clearly in the \feii\ frame (L20-105).    Two additional Fe-detected objects (L20-121 and L20-124) appear to have faint counterparts in the WFC3 optical bands and represent two new SNR candidates.  The red circle indicates the position of an extended ($\sim 1\farcs 3$ diameter) faint L19 optical SNR that has surface brightness too low for detection from the WFC3; it has no discernible \feii\ counterpart.

In Fig.\ \ref{fig_example3} we show a small region with two optical SNR candidates. L19-095 is very compact in {\em HST} optical bands and has a bright \feii\ counterpart (L20-087) that looks larger, probably due to the larger pixels in the WFC3 IR camera.  L19-096 is much fainter but is resolved into a small optical shell.  Interestingly, this object appears to have a comparable \feii\ counterpart that we failed to identify in the blind search. The bright, compact \hii\ region at upper right in this figure highlights once again how any \feii\ from the \hii\ region is much fainter than the \PaB\ and/or \ha\ emission.

\subsection{Sizes and Fluxes \label{sec_size_flux}}

Next we measured the diameters of each object.  Our primary intent was to measure size in the \feii\ images,  but we also inspected the \ha\ and \sii\ images, as there were cases where  higher resolution of WFC3 in UVIS made it easier to select and measure the object in question. The diameter estimates are somewhat subjective, but independent measurements by two of us suggest a consistency of about 0.1\arcsec.  The sources range in diameter from 0\farcs 16 to 1\farcs 3, which corresponds  to 6.0 to 49 pc at our assumed distance of \gal.

We used the size estimates to extract count rates, in ${\rm electrons\:s^{-1}}$, in the drizzled images for the sources in \feii, \PaB, and \ha.  Specifically we summed the counts within a circular (or in a few cases elliptical) region, and subtracted a local background measured in an annulus surrounding each object.  To guard against outliers in the background region, we used the median value derived from each background region to perform the subtraction.  We have estimated the uncertainties based on the standard deviation of the counts in the background region for each object, but note that the results are quite sensitive to the background estimate and the uncertainties for large, low-surface-brightness objects may be somewhat larger than the statistical errors suggest.   An additional concern is that some of the \feii\ objects, flagged as being in complex environments in Table~\ref{ngc6946_tab1}, are contained within larger nebulae visible in \PaB\  and/or  \ha.  We have used the size of the \feii\ object to determine the extraction and background region sizes in all of the filters.

The derived counts are converted to fluxes based on {\em HST} Exposure Time Calculator (ETC) estimates.\footnote{The {\em Hubble} Exposure Time Calculator is kept up to date with the actual sensitivities derived from calibration data, and hence provides a convenient way to convert derived counts into fluxes.  See the ETC User Guide for more information: \url{http://etc.stsci.edu/etcstatic/users_guide/index.html.}}  {In some cases, the net counts in the \PaB\ filter were negative because the average count rate in the source region was less than that of the background region; these cases are responsible for the negative flux values reported in Table~\ref{ngc6946_tab1}.}  The \feii\ fluxes of the objects range from \EXPU{4}{-17} to \EXPU{6}{-15}{\FLUX}, corresponding to luminosities of \EXPU{3}{35} - \EXPU{4}{37}{\LUM}.    The brightest \feii\  source by far is L20-094 = L19-098, originally identified as a SNR by \cite{blair94}, but later associated with an ultraluminous X-ray source (ULX) by \citet{kaaret10}. This object is discussed in more detail in Sec. \ref{sec_histSNe}. 

\subsection{The source catalog}

All of these results are tabulated  in Table \ref{ngc6946_tab1}, which  includes an object name, the position of the object, its apparent diameter (converted to pc), its galactocentric distance, the measured fluxes in \FeiiL\ and \PaB,  and any coincidences with optical SNRs identified by L19, radio SNR candidates identified by \cite{lacey01}, X-ray sources identified by \cite{fridriksson08}, and/or \feii\ objects identified by \cite{bruursema14}, in their previous ground-based  search.

\section{Analysis\label{sec_analysis}}

\subsection{Comparison to the Optical SNRs in L19}

Of the 147 optical SNRs and SNR candidates identified by \cite{long19}, 92 are in the field surveyed in \feii\@.  Of these, 54 (nearly 60\%) are among our list of \feii\ emission nebulae.  The L19 list  included all of the 27 emission nebulae that had been previously identified as SNR candidates by \cite{matonick97}; of the 19 of these that were within the \feii\ observation footprint, we found 15 (nearly 80\%) in the  \feii\ blind search.\footnote{The higher percentage of the \cite{matonick97} sources detected simply reflects the fact that the SNRs detected by \cite{matonick97}  are on average brighter than those in the more sensitive survey conducted by L19 more than 20 years later.}


 There are an additional 38 L19 SNRs within our \feii\ survey region that were not detected in the blind search.  We looked  visually at these source positions and found that a significant number have real, or at least plausible, \feii\ emission coincident with the optical SNR, but at such a low level  that they were not identified in our blind search. 
 
\begin{figure}
\plotone{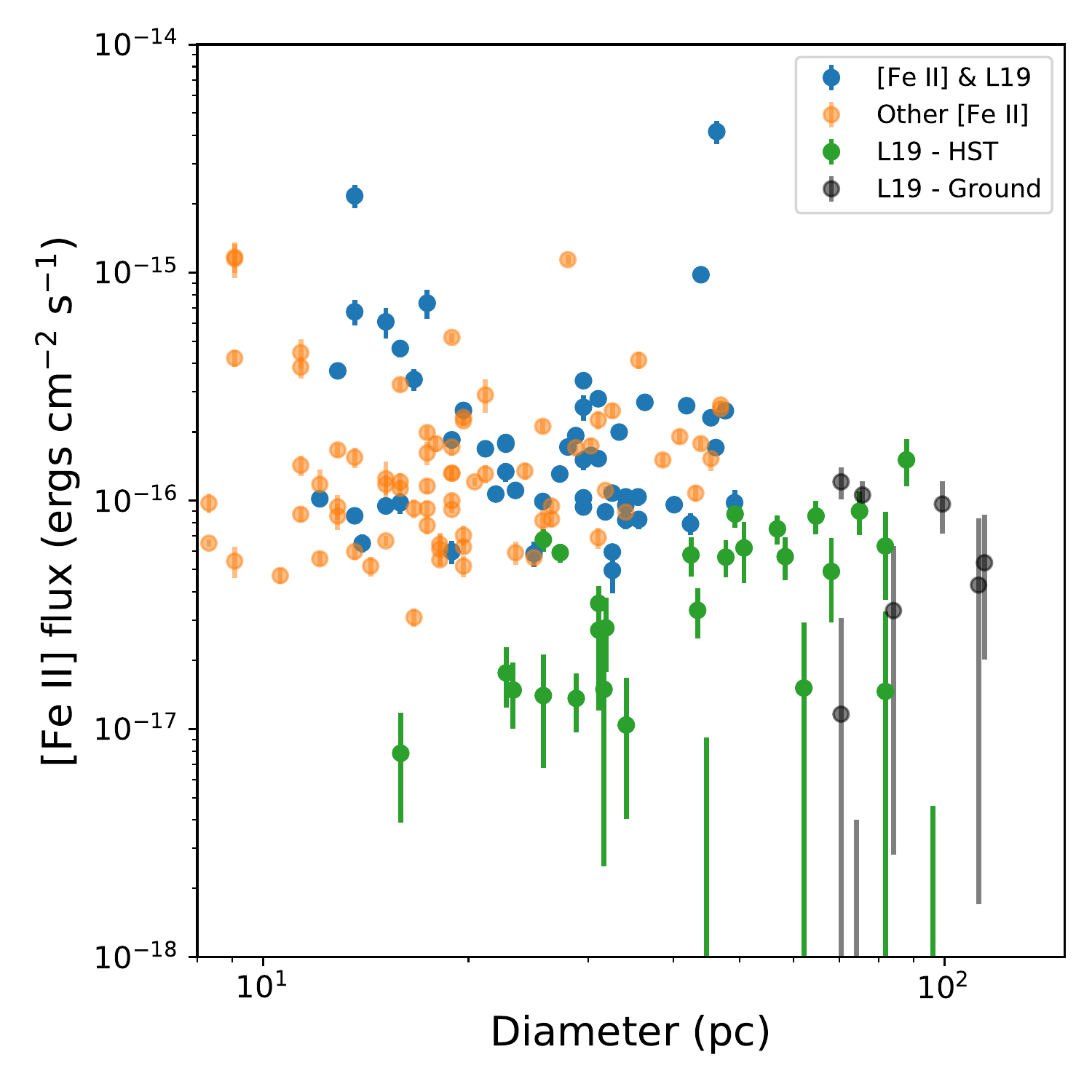}
\caption{\feii\ flux, with 1$\sigma$ statistical uncertainties, as a function of diameter.  Objects found in our blind search that had previously been identified as SNRs by L19 are shown in blue, while those that do not appear in L19 are shown in orange.  \feii\ fluxes extracted at the positions of SNRs  seen only in L19 are plotted in green if we were able to establish an diameter from the HST images,  and in brown if we used  ground-based images (with poorer resolution) to measure the diameter.  The dispersion in luminosities at any given diameter is quite large, and there is little correlation between flux and diameter.  \label{fig_flux}
}
\end{figure}

 In an attempt to make our visual inspection more quantitative, we extracted \feii\ fluxes at the positions of all L19 optical sources, using sizes measured from {\em HST} images where they were available, and from ground-based images where they were not.
These fluxes, along with the fluxes of the emission nebulae found in our blind search,  are given in  Tables \ref{ngc6946_tab1} and \ref{ngc6946_tab2},  and are plotted as a function of diameter in  Fig.\  \ref{fig_flux}.  
 Of the  38 L19 SNRs not found in the blind search, 19 appear to have been detected at $\ge$3$\sigma$, and 10 were detected at  $\ge$5$\sigma$.    These are {\em not} included in Table~\ref{ngc6946_tab1}, since the goal of the present study has been to assess the utility of narrow-band \FeiiL\ images as independent probe to identify SNRs.  Sources that were not found in the blind search tend to be larger diameter, low-surface-brightness objects, which made them harder to pick out by visual inspection.    Also evident from the figure is that there are no obvious trends in the fluxes as a function of diameter, except perhaps a slight trend toward lower luminosities at larger diameter.  The dispersion in luminosities at any particular diameter is quite large.   
 
 If we take 5$\sigma$ as the threshold, 64 of the 92  optical SNRs (70\%) were likely detected in \feii\ at some level. The detection of these objects provides more support that these emission nebulae are SNRs, and further strengthens the argument that \feii\ can be an efficient way to find SNRs in nearby galaxies.

\subsection{Association with known sources in other surveys}

With our list of \feii\ sources from the blind search in hand, we looked for coincidences with objects that had been suggested as SNRs previously.
We can also compare directly to X-ray sources identified in the {\em Chandra} data \citep{fridriksson08}. In addition to the more complete discussion of these below, the coincidences are summarized in Table~\ref{tab:overlap}. Note that the radio source list of interest is from \citet{lacey01} but the source numbering of the radio sources is from the original survey of \citet[][denoted as L97]{lacey97}.

One of our primary motivations was to see whether the sources identified as radio SNR candidates by \cite{lacey01}, only seven of which had optical counterparts in L19, would be detected in \feii.  Of the 35 objects \cite{lacey01} identified as radio SNR candidates based on non-thermal radio spectral indices, 30 were in the field of our \feii\ survey, and 13 have coincident \feii\ emission nebulae detected in the blind search---including one (L97-026) with two bright and clearly distinct \feii\ sources, both within 1\arcsec\ of the radio position (L20-020 and L20-022).  One radio source, L97-34, aligns with the optical SNR L19-066, for which  \feii\ was not found in the blind search.  However, Table 1 shows that a statistically significant \feii\ detection was made in the quantitative assessment, so we include this source among the radio detections, resulting in 14 of 30 radio sources detected in \feii\@. Obviously, this represents a substantial increase, but there are still 16 radio sources within the \feii\ footprint for which no optical or \feii\ source has been identified.  

\cite{lacey01} had argued that the reason more of the radio SNRs in nearby galaxies (and in  \gal\ in particular) had not been detected optically as SNRs  was that radio SNRs were concentrated in the spiral arms, which made detection against a background of dust and \ha\ emission more difficult.  The fact that we are now finding more of these objects in \gal\, using \feii\ imagery, which is less sensitive to both dust and \hii\ region emission, seems to validate that argument.

SNRs are also often detected as soft X-ray sources, although in general X-ray surveys have not been sufficiently deep to find SNRs in galaxies beyond the Local Group via their X-ray emission.  In the case of \gal,
the large foreground  column density from its position at low galactic latitude likely impacts the soft X-ray emission normally seen from SNRs, so that only the brightest SNRs are expected to be detectable.
The most detailed X-ray study of \gal\ to date has been carried out by \cite{fridriksson08}, who identified 90 discrete X-ray sources. There are 67 X-ray sources in the field of the \feii\ images, but there are only 15 coincidences with \feii\ nebulae.  There are six coincidences between X-ray sources and the \cite{lacey01} radio SNRs;  five of these are strong \feii\ sources listed in Table~\ref{ngc6946_tab1}, including L97-026 = F08-23 with two coincident \feii\ sources, as noted above.  Furthermore, careful examination of the position of the sixth radio-X-ray coincidence position (L97-043 = F08-38)  shows that it too has a small, faint \feii\ source.  As with the 11 faint \feii\ sources associated with L19 optical SNRs, we have not included this source in Table~\ref{ngc6946_tab1}. 

There are five remaining X-ray-\feii\ coincidences for which there are no previous SNR candidates at radio or optical wavelengths.  Two of these five, L20-067 and L20-105, have compact emission nebulae in the {\em HST} optical data that strengthen their identification as SNRs. One of the objects, L20-131, is projected within a complex \hii\ region, so it is not clear if the X-ray emission arises from the \feii\ source or from something else.  Finally, there are two X-ray-\feii\ sources, L20-070 and L20-084, that were identified as overlaps, but for which the alignment is not as good as the other coincidences.  Here too, it is not clear if these two are real associations.

We also compared our \feii\ source list to the catalog of objects identified by \cite{bruursema14} as potential \feii\ nebulae based on ground-based imaging.  Of the 48 candidates they suggested, 33 were in our search field, but only six of these were detected---all of them among the brightest objects in our survey.  Given the higher spatial resolution and greater sensitivity of the {\em HST} observations, it seems likely that most of the objects identified by \cite{bruursema14} must have been artifacts or noise peaks in the ground-based data, something they noted as a possibility at the time.  This appears to be confirmed by more recent ground-based work as well (Bruursema, private communication).

\subsection{Completeness of the \feii\ Catalog \label{sec_complete}}

{The completeness of the \feii\ catalog is not straightforward to estimate.\footnote{In principle, one could conduct artificial ``star'' tests to address the completeness if necessary, but the significant effort to do this does not seem warranted.}  As noted in Sec.\ \ref{sec_size_flux}, the diameters of the \feii\ nebulae range from 6 pc to 49 pc.  There was nothing about our search technique that discriminates greatly against objects which are either smaller than 6 pc or larger than 49 pc if they are luminous enough. We did not place an artificial limit on source size.    Since the point spread function of the \feii\ images corresponds to 5.7 pc at a distance of 7.8 Mpc, and  since we used a visual estimate of the source size, we would not measure a diameter less than about this, unless the object was visible in the UVIS channel as well.  At large diameters, for fixed total flux, and assuming a relatively smooth distribution of emission, surface brightness considerations limit the objects that can be picked out of the background (particularly if the region has a spatially variable underlying stellar continuum).  

An examination of the distribution of fluxes in various size ranges indicates that the catalog is better described as flux-limited rather than surface brightness-limited.  Comparing the 30 catalog objects with diameters less than 15 pc to the 37 objects with diameters greater than 25 pc, we find that the median (minimum) fluxes are \EXPU{1.0}{-16} (\EXPU{4.7}{-17}) $\FLUX$ and  \EXPU{1.5}{-16} (\EXPU{5.0}{-17}) $\FLUX$, respectively, implying that large objects are only slightly more luminous than small ones. By contrast the median (minimum) surface brightness is nearly an order of magnitude lower for large objects \EXPU{1.9}{-15} (\EXPU{4.5}{-16})  $\FLUXARCSEC$ versus  \EXPU{1.9}{-16} (\EXPU{7.4}{-17}) $\FLUXARCSEC$, respectively.

Of the 132 nebulae in the catalog, 90\% have \feii\ fluxes exceeding \EXPU{6}{-17}{\FLUX}.  
However, at that limit there will be some \feii\ nebulae that we will have missed in \gal. }
 First,  the  background  stellar density varies considerably  across  the  face  of  the galaxy,  especially  toward  the  nucleus. Thus, the quality of the continuum subtraction varies from one place to another.  Second, color differences of individual sources within the fairly broad bandpass of our continuum filter means that there is no single scaling of continuum to emission-line images that produces a perfect subtraction for all stars.  We chose a continuum scaling that appeared to provide the best overall subtraction of the emission-line scene, with some stars over-subtracting and others under-subtracting. Since most  of  the  SNRs  in  \gal\ have diameters of significantly less than 1\arcsec, it can be difficult to distinguish them from  stellar residuals in the {\em HST} images.  The extremes of over- or under-subtracted are easy to see in the resulting subtracted scene, but there are many intermediate cases that are harder to diagnose with certainty.  And third, many of the faintest optical SNR candidates have no identified \feii\ counterpart.  

\subsection{Historical SNe and Other Objects of Interest\label{sec_histSNe}}

\gal\ has hosted ten SNe within the past century, but several of them, including SN 1980K, SN 2004et, and SN 1939C\footnote{The nominal SN 1939C position very nearly coincides with a bright foreground star that is unrelated to NGC\,6946.}, are outside the coverage of our {\em HST} data.  Also, the recent ``failed SN" that apparently collapsed directly into a black hole \citep{murphy18}, NGC\,6946-BH1, is outside of the current data coverage.

We have examined the {\em HST} images at the positions of the seven remaining historical events, and the only object that shows anything of interest in either our \feii\ or UVIS data is the most recent object, SN 2017eaw, which was a type II-P SN with a red supergiant progenitor \citep{vandyk19}.  Late-time spectra \citep{szalai19} showed the object to be in the nebular phase, and so its presence as an unresolved \ha\ and continuum source in our UVIS data from 2019 January, some $\sim$570 days post-explosion, is perhaps not unexpected. Although there appears to be a faint \feii\  source at the position, this may simply be a subtraction residual from the continuum source.

Another source of particular interest is the ultraluminous X-ray source NGC\,6946-ULX, L20-094, which was originally detected optically by \citet{blair94} and appears as object 16 in the SNR list of MF97, and as L19-098.  This source has been observed previously with {\em HST} \citep{dunne00,blair01}, and was originally thought to be an extremely luminous SNR\@.  However, the properties of the X-ray source and discovery of rapid X-ray variability \citep{roberts03,rao10} make it clear that a compact accreting source is also involved. \citet{roberts03} claim the {\em Chandra} X-ray source is consistent with a point source, but both optical \ha-\sii\ imagery and spectra indicate an extended structure with high \sii:\ha\ ratio indicative of shock heating.

We show the L20-094 region from our new {\em HST} data in Fig.\,\ref{fig_mf16}.   The optical emission in \ha\ and \sii\ in the current data have higher signal-to-noise than previous images, but show essentially the same structure: a small, bright shell to the west side and multiple shells or loops extending more than 1\arcsec\ to the east. The object has not been imaged previously in \feii\ and \PaB, and the WFC3-IR data show much the same structure as seen at optical wavelengths (but at slightly lower resolution due to the larger pixel size of the IR camera). In addition to bright \feii, Pa$\beta$ emission, consistent in extent with \ha, is seen, but the observed \feii:Pa$\beta$ ratio of 7.0 (see Table 2) lies clearly in the  shock regime. 
The small shell on the west side is very bright in \feii\ and likely dominates this measurement.  However, by assessing the relative counts in the {\em HST} data, the extended structure to the east is clearly seen in \feii\ as well, with \feii:Pa$\beta$ $\simeq$ 2.3, confirming the shocked nature of this extended emission.    The stellar source that lies within the extended structure (and visible in the F547M data) has been noted previously \citep{blair01}.

\section{Discussion \label{sec_discussion}}

\subsection{Theoretical Expectations\label{sec_theory}}

From comparing the strength of \feii\ emission at the location of optically-identified \hii\ regions and SNRs,  it is clear that the ratio of \feii\,:\,\PaB\ (or some other hydrogen line) is much higher in the shock-heated objects.  To first order, this is expected, for the same reasons that \sii:\ha\ is high in shocks compared with \hii\ regions. The ionization potential for $\rm Fe^{+}$ is only 16.2 eV, and so in photoionized regions most of the Fe has been ionized to Fe$^{++}$ or above,  and hence \feii\ emission is relatively weak.  In fully developed shocks, there is a cooling and recombination zone where the $\rm S^{+}$ and/or $\rm Fe^{+}$ ions dominate, and hence these low-ionization lines are relatively strong. 

To place this expectation on firmer theoretical ground requires model calculations, and the literature is relatively scant in this regard, especially for \hii\ regions.  Perhaps the best overall summary available is the paper by \citet{mouri00}, who used the {\em Mappings III} code \citep{dopita96} to study the expected \feii\ emission from shocks as well as from both blackbody and X-ray photoionization (the latter being appropriate for Seyfert galaxies and AGN, which can also be significant \feii\ sources).  \citet{mouri00} concentrate on comparing the ratio of \OiL\,:\,\ha\ with \FeiiLL\,:\,\PaB\ (note that \FeiiLL\ is the other strong near-IR iron line).  The trend in \feii\,:\,\PaB\ is clear: for typical \hii\ region and SNR densities and typical (iron-depleted) ISM  abundances, \feii\,:\,\PaB\ is high in both shock models and X-ray photoionization models compared with normal blackbody photoionization.  \citet{mouri00} show results from shocks in the range of 50 -- 150 $\VEL$, which are typical of bright radiative SNRs, but do not cover the entire expected range of parameter space for SNR shocks.  The key to enhanced \feii\ emission, according to these authors, is the presence of an extensive partially-ionized zone (or in the case of shocks, a recombining plasma zone) that is simply not present in regular \hii\ regions.

\citet{allen08} provide a much broader grid of {\em Mappings III} shock models.  While the figures in that paper do not include the \feii\ lines, they are included in the online library of those models. Also, \citet{koo16} have used the updated shock code of \cite{raymond79} to explicitly investigate \feii\ emission from shocks.  The \citet{allen08} models cover an extensive grid of shock velocity and other parameters.  While the \feii\,:\,\PaB\ ratio is not universally high over this full grid, over the range of most likely velocities for typical radiative SNR shocks, the ratio is indeed high. \citet{koo16} concentrate on slow to moderate shock velocities (20 -- 200 $\VEL$) and predict \FeiiL\ comparable to or stronger than \hb\ over this entire range (see their Figure 7 and Sec.\ 3.3 of their paper), and hence, the expected  \FeiiL\,:\,\PaB\ ratio is much higher than unity.  While \citet{koo16} note some differences in comparing their models to {\em Mappings III},  the sense of these calculations makes it clear that elevated \feii\,:\,\PaB\ emission, such as what we observe for many sources in \gal, is a definite indication of shock heating.

\subsection{Are all of the [Fe~II] sources SNR candidates?}

Of the 132 \feii\ sources identified in our blind search, we found 54 sources that overlapped with L19 optical SNRs, so we consider these sources to be {\em bona fide} SNRs.  As discussed in Sec.~\ref{sec_catalog}, we
have compared the \feii\ source positions of the 78 additional \feii\ sources with the {\em HST} optical data obtained under the {\em HST} program 15216, and have found that 44 of these sources  align with previously unknown optical SNR candidates that  are visible in \ha\ and \sii\ in the new {\em HST} images. Hence, between the L19 overlaps and our new assessment of {\em HST} optical data, a total of 98 out of the 132 \feii\ sources (74\%) have identifiable optical counterparts.
The other 34 sources are seen in projection against complex regions of \hii\ emission and/or dust lanes intrinsic to the spiral arms on NGC\,6946.  We now look at some examples of these additional \feii\ sources to gain insight into their viability as SNR candidates.

\begin{figure}
\plotone{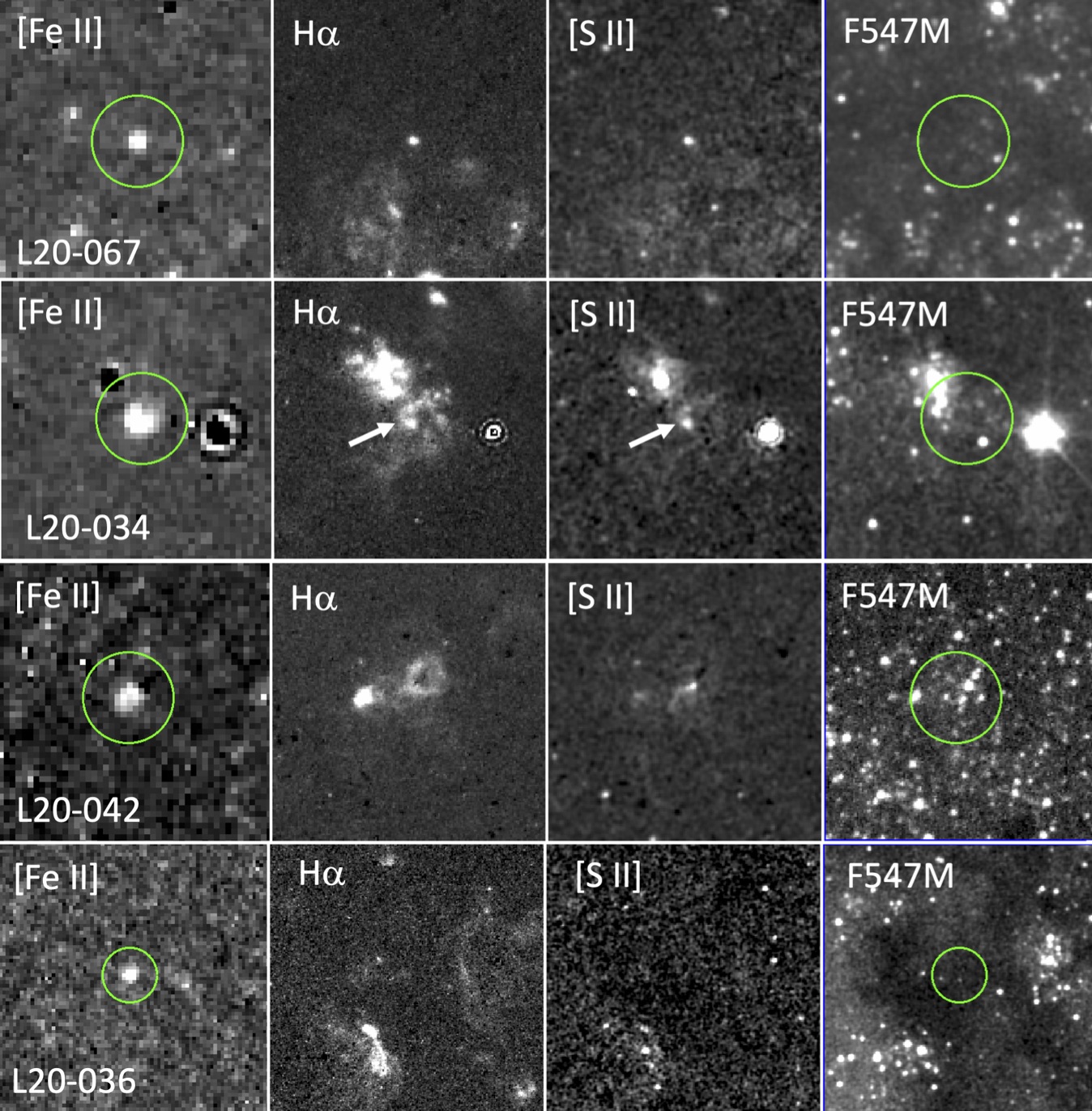}
\caption{Four representative examples of sources from our \feii\ blind search that did not have L19 SNR identifications. Each 4-panel figure shows {\em HST} WFC3 data from \feii, \ha, and \sii\ (all continuum subtracted) and the F547M continuum data for comparison. The green circles are 2\arcsec\ in diameter;  the fields for the upper  three examples are all 6\arcsec\ square, while that for the lower one is 10\arcsec.  The top example shows L20-067, a new SNR whose small angular size and relative faintness made it undetectable in the ground-based SNR search despite the fact that it is well isolated from confusing emission.  Next, L20-034 shows a new SNR  in a complicated region of optical emission.  At {\em HST} resolution, the optical SNR can be seen as a partial shell with a bright knot of emission toward the south (arrows), but in \feii\ the object stands out clearly, without the confusion seen in \ha\ and \sii\@.   Not surprisingly, this object also was  not detected in  ground-based data.  The third panel shows L20-042, an \feii\ source loosely associated with \hii\ emission but with no obvious optical SNR candidate.  If this is indeed a SNR, it must be behind dust and/or simply have faint optical emission.
Finally, L20-036 shows a well-detected \feii\ source that has no optical \ha\ or \sii\ counterpart.  However, referencing the F547M frame, the position is projected toward a dark, dusty region. For this and other similar objects, we claim these objects to be SNRs whose optical emission is extincted but whose \feii\ emission gets through the dust.  
\label{fig_new_SNRs}}
\end{figure}

\begin{figure}
\plotone{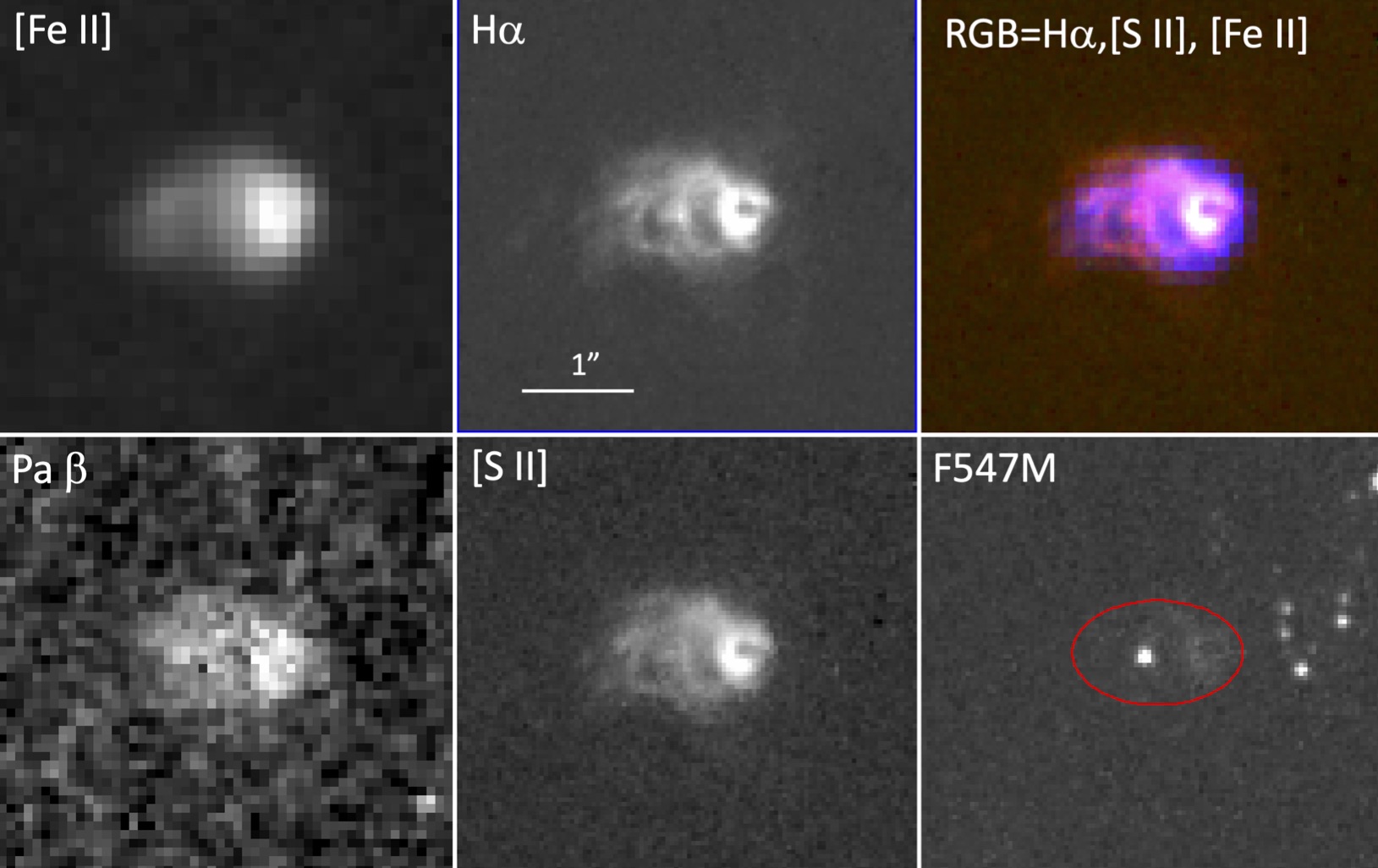}
\caption{This figure shows a close-up view of the NGC\,6946 ULX, also known as MF16 \citep{matonick97} in the previous literature; it is Fe-20-094 in the catalog of Table 1.  The panels show {\em HST} WFC3 IR and UVIS data as labeled and the scale is shown in the top middle panel. As with its optical line emission, IR emission from this object is consistent with shock heating, with \feii\,:\,\PaB\ ratio $\approx 7$.
\label{fig_mf16}
}
\end{figure}

In Fig.~\ref{fig_new_SNRs}, we show four representative examples of the categories of sources we see beyond the L19 SNRs.
L20-067 (Fig.\ \ref{fig_new_SNRs} top) has an isolated, very compact but reasonably well-detected \ha\ and \sii\ nebula aligned with the \feii\ source.  There is no nearby star that could be causing a stellar residual at this location, and we conclude this is a new SNR candidate that was unseen in the ground-based survey.  Very compact sources such as this are resolved by {\em HST} but would be smeared out at ground-based resolution.  
There are a number of other examples both brighter and fainter than this one, down to the limit of detectability in the optical survey.

The object L20-034 is shown in the second row of Fig.\ \ref{fig_new_SNRs}.  In this case, a compact knot and faint partial shell of \ha\ with comparable \sii\ can be seen as indicated by the arrow, even though it resides in a region of complex \hii\ emission.  In the display, the contrast has been adjusted to best show the optical SNR candidate, but significant \hii\ emission overlies the entire region.  No stellar source is present at this position, although a number of bright stars are nearby, exciting the \hii\ gas.  It is easy to see why such a source would be missed in ground-based data, but the \feii\ source is  bright and easily detected.  Again, there are a number of similar sources identified by comparing the \feii\ source positions to the optical data.  In more extreme cases, there can be a clear \feii\ source seen in projection against \hii\ emission but where we cannot identify a specific optical counterpart, as shown in the third row of the figure (L20-042).  It is hard to imagine such a source being anything other than a SNR, perhaps behind enough local extinction in the \hii\ region that the optical counterpart is hidden.

The final example is L20-036, shown in the bottom row of Fig.\ \ref{fig_new_SNRs}.  In this case, a well detected \feii\ source has no counterpart at all in the optical data, but the source is projected against a dark region of interstellar dust, as seen in F547M\@. It seems likely that this is a SNR that is behind the dust cloud, so that its optical emission is too heavily extincted  to be detected, but its  less absorbed  \feii\ emission  is able to penetrate the dust.  We find a number of intermediate cases of very faint optical SNR candidates that align with \feii\ sources that are likely at varying depths into dusty regions.

From this inspection, we are left with the impression that all of the sources in our \feii\ blind search list are  strong SNR candidates.  
If we had searched the optical {\em HST} data separately prior to comparing with the \feii\ source list, presumably a number of these objects would have been identified independently.   Many of the other objects 
are seen much more readily in \feii, where the \hii\ emission essentially disappears.  It seems likely then  that the 34 \feii\ sources without optical counterparts represent SNRs whose \feii\ emission is visible but for which either dust absorption and/or overlying complex emission prevents the optical SNR from being seen. This is reminiscent of the situation seen previously in M83 \citep{blair14}, and is another strong demonstration of the power of using \feii\ in combination with optical criteria to determine more complete samples of SNRs in nearby galaxies. 

Thus, if we consider all 147 L19 optical SNRs and the 78 additional \feii\ sources discovered here that did not have L19 identifications, there are now some 225 total optical/IR SNR candidates in NGC\,6946.   There are still 21 of the 35 non-thermal radio sources from \citet{lacey01} without optical/IR counterparts that are also in contention as possible SNRs.  Table~\ref{tab:overlap} summarizes all of the multiwavelength overlaps.

\subsection{[Fe~II] versus Pa$\beta$}

Although \feii\ emission is relatively faint in \hii\ regions, it is not expected to be entirely absent.  We do see faint  diffuse \feii\ emission at the positions of some of the brightest \hii\ regions, but that emission is readily separable from the clumps or knots of \feii\ emission we have identified in the blind search list.

\begin{figure}
\plotone{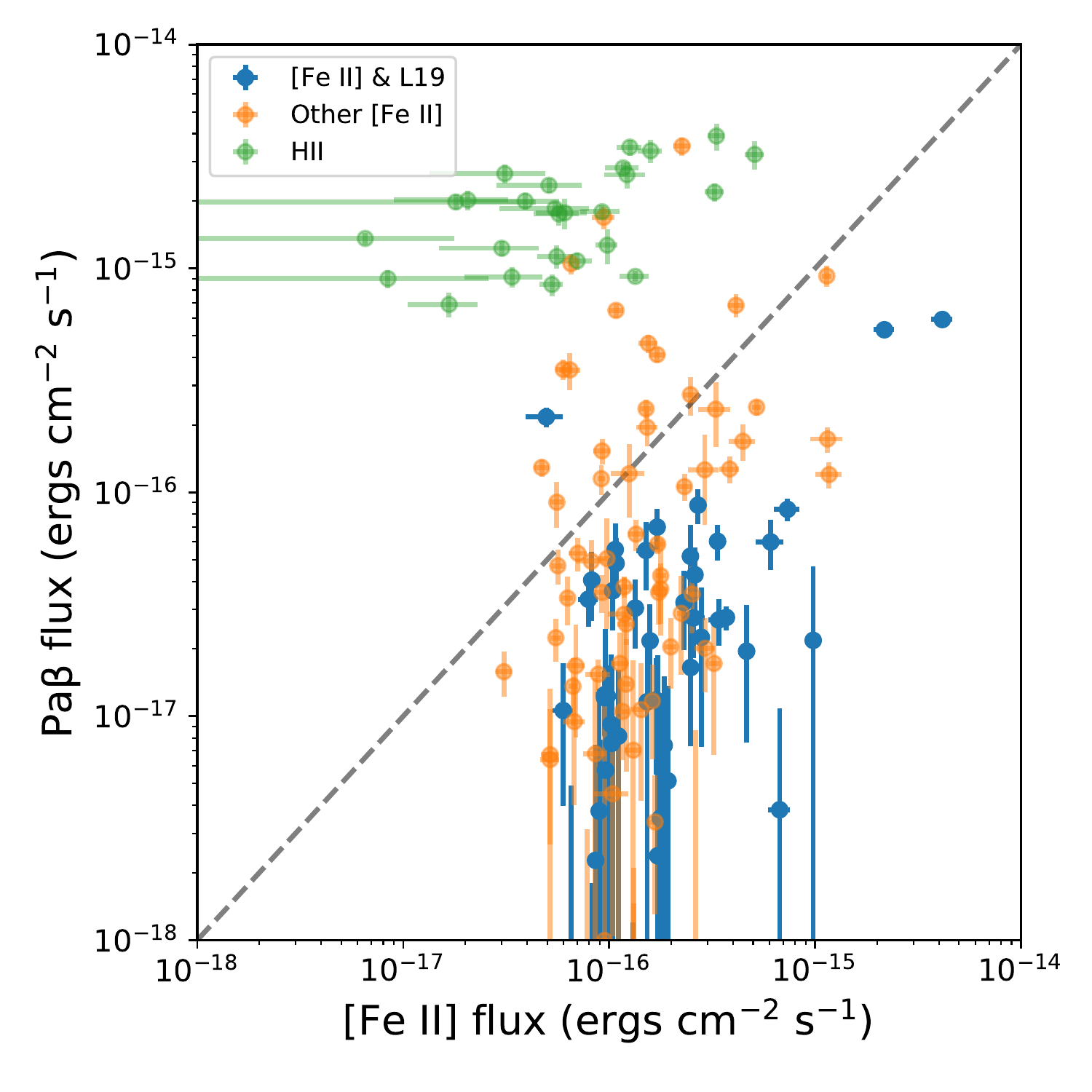}
\caption{Measured fluxes of the \feii\ sources in \feii\ and \PaB, with 1$\sigma$ statistical uncertainties.  Fluxes for a set of \hii\ regions  are also shown. The \feii\ sources that are coincident with optical SNRs have a median \feii:\PaB\ ratio of about 8, while the other \feii\ sources have a median value of about 3.  The objects not identified with optical SNRs but with high \feii\,:\,\PaB\ ratios (i.e. below the dashed line) are likely SNRs. The compact \feii\ sources with with lower ratios (above the dashed line),  even the few amidst the \hii\ region points, are likely to be SNRs buried within varying amounts of overlying \hii\ emission.  
\label{fig_pab_fe2_ratio}}
\end{figure}

Accordingly, we can expect that the ratio of observed  \feii\ to H emission  should provide a diagnostic for shocked versus photoionized gas.\footnote{Throughout this discussion, we equate evidence of shock-heating with SNRs and photoionization heating with HII regions.  There are other objects where \feii\ is observed due to shock heating, such as Herbig Haro objects \citep{nisini02}, luminous blue variables \citep{smith06} and even some planetary nebulae \citep{smith05}.  However, these objects are far too faint to have been detected in \feii\  in \gal. }  Although \ha\ is brighter than \PaB, the \PaB\ line is not contaminated by other emission lines (as is the case for the WFC3 ``\ha" images, where the F657N filter also passes the \NiiL\ lines), and \PaB\ will also be less affected by differential reddening.  Hence, in Fig. ~\ref{fig_pab_fe2_ratio}, we have plotted the \PaB\ flux as a function of the \feii\ flux.  As is apparent from the figure, the \feii:\PaB\  ratio tends to be high for nebulae that had already been identified as optical SNRs using the \sii:\ha\ method.  By contrast, the objects not  previously known to be SNRs have a broader distribution, including a number of objects where the \feii:\PaB\ ratio is near 1.  Of the 54 L19 optical SNRs that are  in our catalog of \feii\ objects and that are also in the field of the \PaB\ images, 45 have \feii:\PaB\ ratios of 3 or higher.  If we take this as a reasonable boundary, then of the 78 \feii\ nebulae not known from L19 and for which we have \PaB\ images, 38 have a ratio greater than 3.  These 38 constitute the strongest candidates to be new SNRs in our sample.

Fig.~\ref{fig_pab_fe2_ratio} also shows a set of points that were derived for {\em bona fide} \hii\ regions in \gal.  For these, the observed \feii:\PaB\ ratios are all significantly less than 1.  This represents an observationally-determined expectation for this ratio in \hii\ regions and is consistent with the theoretical expectations discussed earlier.  If, as we have argued above, all of the compact \feii\ sources we have identified are really SNRs, the \feii\ sources with intermediate values of the ratio can be explained as SNRs within increasing \hii-contaminated sight lines where overlying \PaB\ emission causes the observed ratio to approach the value seen in \hii\ regions.  {The fact that SNRs in the L19 sample tend to have higher  \feii:\PaB\  ratios than the \feii\ nebulae not identified with L19 objects partly reflects the history of the observations.  The optical study was done first and found the bright objects that would have been detected in both cases. However, because the ground-based PSF was larger than the actual diameter of the SNRs in the survey and because the initial visual identification of a nebula as a SNR candidate depended on the \ha\ flux (or more properly the \sii:\ha\ ratio) within the PSF, SNRs embedded in \hii\ regions would have been relatively
hard to pick out unless they were bright (or had intrinsically higher \sii:\ha\ ratios).  This plausibly creates a bias in such regions against detecting candidates with intrinsic  \sii:\ha\ ratios just above the ratio of 0.4 used to define a good candidate, and by extension relatively low \feii:\PaB\  ratios.   In the current {\em HST} survey, the \feii\ and \PaB\ fluxes are typically extracted from smaller regions, so we were able to detect small, shock-excited SNRs despite their locations amidst  extensive \hii\ emission.   Although the new \feii\ nebulae do tend to have lower \feii:\PaB\  ratios than those in L19 sample, there is no obvious trend of this ratio with diameter in either the \feii\ catalog objects or the L19 objects.}

\subsection{Size distribution for \gal\ SNRs}

\begin{figure}
\plotone{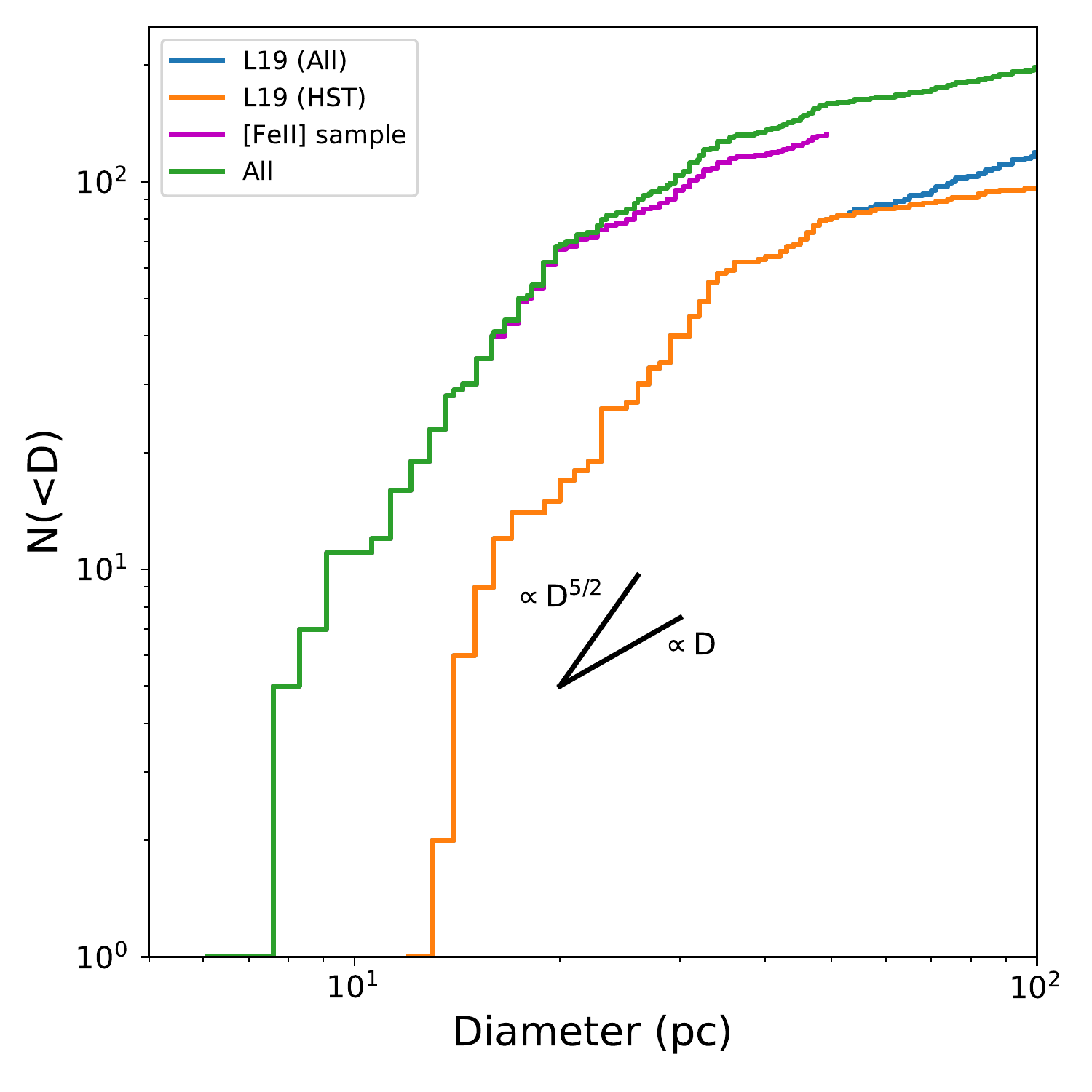}
\caption{The cumulative number of optical and \feii\ SNR candidates  smaller  than a given diameter.  Various subsets of entire sample of SNRs are shown, the entire optical sample from L19 (blue), the portion of the sample with  {\em HST}-measured diameters (orange), and the \feii\ sample (purple),  as well as the entire sample of 225 \feii\ and optical SNRs and  candidates (green).  The slope of the distribution for the large number of SNRs with $10\:{\rm pc} \lesssim D \lesssim$ 20\,pc is consistent with Sedov expansion ($D(t) \propto t^{2/5}$, so $N(<D) \propto D^{5/2}$ for a uniform SN rate), but this may be fortuitous (see text). 
\label{fig_n_diam}
}
\end{figure}

We did not have sufficient angular resolution in the existing ground-based images (L19) to accurately measure diameters for SNRs in our optical sample.  However, the majority of those objects lie within the field of the recent WFC3/UVIS images, which we have now used to measure their diameters. These are provided in Table \ref{ngc6946_tab2},  which also provides the names of the various optical SNR candidates, their positions, and galactocentric distances, as well as an indication of which have been spectroscopically confirmed to have \sii:\ha\ ratios greater than 0.4, and also which are in the \feii\ field of view.  
We find that the \feii-detected objects  are, as a group, systematically smaller than the objects in the L19 optical sample.  
This is shown in the form of an $N(\,<D)\;{\rm vs.}\; D$ diagram in 
Fig.~\ref{fig_n_diam}.  The median diameter of the \feii\ objects is 20 pc, compared to 46 pc for the entire optical sample, or 32 pc for the subset with more accurate sizes measured with {\em HST}\@.  However, the difference may be in part a selection effect, since the L19 objects were all selected from ground-based images, which were relatively insensitive to the smallest objects; e.g., \feii\ sources L20-121 and L20-124 in Fig.~\ref{fig_example2}, which were not detected in L19.

The slope of the distribution shown in Fig.~\ref{fig_n_diam} is consistent with the expectations for Sedov evolution with slope $\propto ~ D^{5/2}$; however, this is most likely fortuitous, since to have a physical meaning we would need to be able to assert that the catalog of SNRs is complete (or more properly that the degree of completeness is straightforward to estimate).  That is unlikely to be true for a variety of reasons, the most prominent being that different SNRs are expanding into very different interstellar environments---some more dense and some less. This is reflected in the large variation in \feii\ flux  at any particular diameter, as  indicated in Fig.\ \ref{fig_flux}; similar variations are seen at other wavelength ranges in other galaxies.  Even if, individually, each SNR was following a Sedov expansion law, the ensemble of SNRs would not be expected to do so.  SNRs expanding into dense environments would be expected to be brighter, but evolve rapidly before fading away, while those expanding into more tenuous media will evolve more slowly and (since the fraction of the explosion energy radiated away is about the same) have lower peak luminosities.  

In Sec.\ \ref{sec_complete}, where we discussed some of the observational issues associated with completeness, we were primarily discussing whether we had found all of the SNRs that were brighter than a certain flux in \feii\@.  A different question of completeness is to ask what fraction of the SNe in NGC\,6496 we have seen.  If we take the ten SNe seen in just over a century as typical, then we would expect $\sim$1,000 SNe in 10,000 years, yet  we have found of order only 20--25\% this many. 

An alternative way to estimate the number of SNe expected is based on the star formation rate, where one expects, according to \cite{maoz17},  0.010$\pm$0.002 CC SNe $M_{\sun}^{-1}$, assuming a \cite{kroupa01} IMF\@. (Here the mass being referred to is the mass participating in star formation.) For a star formation rate of $3.2\, M_\sun\,{\rm yr}^{-1}$  \citep{jarrett13}, this would predict a smaller number, $\sim$320 SNe in 10,000 years.  However, with the relatively recent adjustment in the distance assumed here, \cite{eldridge19} have argued convincingly that the star formation rate from \cite{jarrett13} is too low and is inconsistent with the observed SN rate.  The larger distance compared with that assumed by previous investigators implies that the quantities, such as H$\alpha$ luminosity used to estimate the SFR, are also higher, resulting in a SFR of 12.1$\pm$3.7 $M_\sun\,{\rm yr}^{-1}$.  

Could the last century have produced an inordinate number of SNe?   This would require a serious statistical anomaly, since the number of detected SNRs implies a mean SN rate of only 2-3 SNe per century.  The typical age of the SNRs in our sample must be of order thousands of years. If it were much less, say 1000 years, then L19 should have found some young SNRs with significant line broadening and/or dominated by emission lines from SN ejecta; yet none were seen. Hence, we must still be detecting only a fraction of the actual SNe in our SNR surveys.

\subsection{A Comparison to M83}

There have been very few systematic surveys for \feii\ emission from SNRs in other galaxies.  Much of the initial interest in \feii\  was associated with observations of radio SNRs in galaxies such as M82 \citep{greenhouse91,greenhouse97} and NGC\,253 \citep{forbes93}, but limitations associated with the available detectors  made general surveys difficult.  \cite{morel02} attempted observations of 42 of the SNRs then known in M33, but reported robust detection for  only 9 of them.\footnote{There are about 220 SNR candidates in M33 today \citep{long18}, but unfortunately there has been no more recent study of the NIR \feii\ lines in what is probably the best studied sample of SNRs  in any external galaxy more distant than the Magellanic Clouds.}

The only other galaxy where  \feii\ fluxes for a large number of SNRs have been reported is M83, where \cite{blair14} reported the detection of 51 of 63 small-diameter SNRs using {\em HST}\@.  The fluxes they reported ranged from a minimum of \EXPU{9}{-18} to \EXPU{2.6}{-15}{ergs~cm^{-2}~s^{-1}}, fairly similar to the SNR fluxes we observe in \gal.  The SNRs discussed by \cite{blair14} represent a subset of the SNRs and SNR candidates that have been identified in M83, which currently numbers 304 \citep{williams19}.   To determine what fraction of the entire M83 sample is detected in \feii, we have extracted fluxes from all of the SNR and SNR candidates for which \feii\ data exist.  Of the 304 objects in the sample, there are 262 that lie within the fields covered by the narrow F164N WFC3 filter, and of these  we find that 148 (180) are detected at 5 (3) $\sigma$.  Qualitatively, as in the case of \gal, we find that a larger fraction of small diameter objects is detected.  A systematic search for \feii\ emission nebulae in M83 will be the subject of a future paper.

\section{Summary \label{sec_summary}}

From a pragmatic standpoint, most known SNRs in nearby galaxies have been identified as emission nebulae with \sii:\ha\ ratios in excess of those seen in \hii\ regions. However, due to the diverse, multiwavelength properties of SNRs in general, no single technique is expected to locate the entire population of SNRs.  In order to obtain a more complete picture, it is important to establish alternative criteria for identifying SNRs.  

We have reported here the results from a near-IR survey of the nearby starburst galaxy \gal, carried out through {\em HST} WFC3 images in the \FeiiL\ and \PaB\ emission lines as well as broader continuum filters---complementing our earlier ground-based optical survey \citep{long19}.   
We have shown that the IR line ratio of \FeiiL:\PaB\ appears to be another effective diagnostic for distinguishing SNRs from \hii\ regions, especially in situations where the optical \sii:\ha\ ratio is of limited value due to foreground extinction or overlying complex emission.   

In a sample of  132 \feii\  emission nebulae we have identified in \gal, 54 are coincident with optically identified SNRs from \citet{long19} and 44 more have newly identified optical counterparts.  For most of these, the \feii:\PaB\ ratios are $>$1 which, from comparison with shock models, suggests they are SNRs.  For compact \feii\ sources located within \hii\ regions, the \PaB\ from the \hii\ region can dominate, causing the observed ratio to be lower, but visual inspection makes it clear that the \feii:\PaB\ ratio is elevated at the specific position of the \feii\ source. 

Moreover, 14 of the \feii\ emission nebulae are coincident with radio sources that \cite{lacey01} suggested were SNRs based on their radio properties.  A number of these objects are in areas where the \sii:\ha\ method seems to have failed, either because of foreground dust or because there was too much photoionized emission in the region, as \cite{lacey01} had suggested.  It is a reasonable hypothesis that {\em all} compact sources with \feii\ emission are SNRs.  Under this assumption, there are now 225 optical/IR SNR candidates, plus the remaining 21 \cite{lacey01} radio sources with non-thermal spectral indices that remain as viable SNR candidates.

The NIRCam instrument on the {\em James Webb} Space Telescope includes filters that will allow similar searches to be carried out with much higher sensitivity and better angular resolution than with {\em HST} and WFC3.  NIRSpec and MIRI could be used to obtain NIR/MIR spectra that may lead to a better understanding of how IR and optical properties of SNRs in nearby galaxies are related.  We look forward to seeing the results of those searches.  For \gal\, in particular, we hope to publish the results of a deeper JVLA survey of \gal\  and a complete analysis of the WFC3 UVIS data in the not-too-distant future.  Deeper X-ray observations with {\it Chandra} would allow a better characterization of the X-ray properties of the SNRs. A more complete spectroscopic study of the SNRS in \gal, especially one with  higher velocity resolution,  could potentially resolve why some SNRs are brighter in \ha\ and [S II], while  others are brighter in \feii\@.  A complete spectroscopic survey could also improve the fidelity of the sample, eliminating nebulae that have crept into the sample, but are really not SNRs.   These kinds of detailed studies are required in \gal\ and in other galaxies, in order to seriously address the central question of SNR research in nearby galaxies, viz.,  to establish what properties of SNe and their surrounding environments determine the observational appearance of SNRs.

\acknowledgments

We thank Jennifer Mack for her advice and help on defining an effective procedure to produce the mosaicked images of all of the data from various UVIS and IR filters used in this paper. Partial support for the analysis of the data was provided by NASA through grant number HST-GO-14638 and HST-G0-15216 from the Space Telescope Science Institute, which is operated by AURA, Inc., under NASA contract NAS 5-26555. PFW acknowledges additional support from the NSF through grant AST-1714281.  WPB acknowledges support from HST-GO-14638-B and HST-GO-15216-A, both to Johns Hopkins University, and for partial support from the JHU Center for Astrophysical Sciences.  

\vspace{5mm}
\facilities{HST(WFC3)}

\software{astropy \citep{astropy},  AstroDrizzle \citep{AstroDrizzle}, matplotlib \citep{hunter07}, DS9 \citep{ds9}}

\pagebreak

\bibliographystyle{aasjournal}

\bibliography{snr}

\clearpage



 \begin{longrotatetable}
\begin{deluxetable}{rccrrrrccccc}
\tablecaption{[Fe II] sources}
\tablehead{
\colhead{Name} &
\colhead{R.A.} &
\colhead{Decl.} &
\colhead{D} &
\colhead{R} &
\colhead{[Fe II] flux$^a$} &
\colhead{Pa$\beta$ flux$^a$} &
\colhead{Env.$^{b}$} &
\colhead{Optical} &
\colhead{Radio$^{c,d}$} &
\colhead{Bruursema} &
\colhead{X-ray$^{d}$}
\\
\colhead{} &
\colhead{(J2000)} &
\colhead{(J2000)} &
\colhead{(pc)} &
\colhead{(kpc)} &
\colhead{} &
\colhead{} &
\colhead{} &
\colhead{} &
\colhead{} &
\colhead{} &
\colhead{}
}
\startdata
L20-001 & 20:34:23.38 & 60:08:18.4 & 41 & 7.3 & 26.1$\pm$1.4 & 4.3$\pm$1.4 & - & L19-014 & - & - & - \\ 
L20-002 & 20:34:28.88 & 60:07:45.2 & 32 & 6.4 & 4.9$\pm$1.0 & 21.7$\pm$2.2 & C & L19-028 & - & - & - \\ 
L20-003 & 20:34:31.06 & 60:08:27.3 & 32 & 5.4 & 24.8$\pm$1.8 & 27.3$\pm$5.3 & C & - & L97-08 & - & - \\ 
L20-004 & 20:34:31.64 & 60:10:27.8 & 35 & 6.2 & 10.4$\pm$0.9 & -0.3$\pm$1.2 & - & L19-031 & - & - & - \\ 
L20-005 & 20:34:32.41 & 60:10:12.3 & 30 & 5.7 & 17.4$\pm$1.4 & 3.6$\pm$1.0 & - & - & - & - & - \\ 
L20-006 & 20:34:32.59 & 60:10:27.7 & 22 & 6.0 & 18.0$\pm$1.6 & -0.8$\pm$0.8 & - & L19-032 & - & - & - \\ 
L20-007 & 20:34:33.14 & 60:10:32.4 & 23 & 6.0 & 5.9$\pm$0.7 & -1.3$\pm$0.6 & - & - & - & - & - \\ 
L20-008 & 20:34:33.29 & 60:09:46.5 & 28 & 5.1 & 19.3$\pm$1.4 & 0.5$\pm$0.8 & - & L19-034 & - & - & - \\ 
L20-009 & 20:34:33.62 & 60:09:51.9 & 31 & 5.1 & 15.3$\pm$1.1 & 1.2$\pm$1.1 & - & L19-035 & - & - & - \\ 
L20-010 & 20:34:36.13 & 60:08:39.2 & 11 & 4.1 & 44.6$\pm$6.6 & 16.9$\pm$3.1 & - & - & L97-17 & - & F08-15 \\ 
L20-011 & 20:34:37.28 & 60:08:37.0 & 31 & 3.9 & 22.6$\pm$2.0 & 352.0$\pm$32.7 & C & - & - & - & - \\ 
L20-012 & 20:34:37.42 & 60:11:31.3 & 40 & 6.8 & 9.6$\pm$0.7 & 0.6$\pm$1.1 & - & L19-039 & - & - & - \\ 
L20-013 & 20:34:37.51 & 60:09:37.0 & 35 & 3.9 & 41.3$\pm$3.6 & 68.4$\pm$8.1 & C & - & L97-22 & - & - \\ 
L20-014 & 20:34:37.76 & 60:08:14.8 & 16 & 4.0 & 9.2$\pm$0.9 & 15.3$\pm$2.0 & C & - & - & - & - \\ 
L20-015 & 20:34:37.96 & 60:07:22.0 & 30 & 5.2 & 15.8$\pm$1.0 & 2.2$\pm$1.0 & - & L19-042 & - & - & - \\ 
L20-016 & 20:34:39.11 & 60:09:18.5 & 23 & 3.3 & 11.1$\pm$0.6 & 0.8$\pm$0.8 & - & L19-045 & - & - & - \\ 
L20-017 & 20:34:39.13 & 60:09:09.6 & 45 & 3.3 & 15.3$\pm$1.8 & 19.5$\pm$3.4 & C & - & - & - & - \\ 
L20-018 & 20:34:39.16 & 60:08:13.7 & 18 & 3.7 & 18.5$\pm$1.6 & 0.7$\pm$0.8 & - & L19-046 & - & - & - \\ 
L20-019 & 20:34:40.67 & 60:06:53.1 & 49 & 5.7 & 9.8$\pm$1.4 & 0.1$\pm$1.5 & - & L19-048 & - & - & - \\ 
L20-020 & 20:34:41.25 & 60:08:46.3 & 11 & 2.8 & 38.5$\pm$4.2 & 12.7$\pm$1.7 & C & - & L97-26 & - & - \\ 
L20-021 & 20:34:41.31 & 60:11:12.6 & 42 & 5.5 & 7.9$\pm$0.9 & 3.3$\pm$0.8 & - & L19-051 & - & - & - \\ 
L20-022 & 20:34:41.42 & 60:08:46.3 & 9 & 2.8 & 117.0$\pm$17.1 & 12.0$\pm$1.6 & C & - & L97-26 & B14-14 & F08-23 \\ 
L20-023 & 20:34:41.56 & 60:06:17.8 & 18 & 6.8 & 10.0$\pm$0.7 & -- & - & - & - & - & - \\ 
L20-024 & 20:34:43.65 & 60:09:10.7 & 15 & 2.1 & 6.7$\pm$0.3 & 1.4$\pm$0.5 & - & - & - & - & - \\ 
L20-025 & 20:34:43.97 & 60:08:24.0 & 31 & 2.6 & 8.9$\pm$0.8 & 0.4$\pm$0.9 & - & L19-059 & - & - & - \\ 
L20-026 & 20:34:44.59 & 60:08:17.0 & 13 & 2.7 & 8.6$\pm$0.5 & 0.2$\pm$0.5 & - & L19-060 & - & - & - \\ 
L20-027 & 20:34:46.68 & 60:10:05.0 & 25 & 2.5 & 5.6$\pm$0.4 & 4.7$\pm$0.8 & - & - & - & - & - \\ 
L20-028 & 20:34:47.37 & 60:08:22.3 & 29 & 2.1 & 25.7$\pm$3.3 & 2.8$\pm$0.9 & - & L19-065 & - & - & - \\ 
L20-029 & 20:34:47.79 & 60:10:51.2 & 31 & 3.9 & 6.9$\pm$0.7 & 1.7$\pm$0.9 & - & - & - & - & - \\ 
L20-030 & 20:34:47.82 & 60:07:51.5 & 19 & 3.1 & 6.3$\pm$0.4 & 3.4$\pm$0.8 & C & - & - & - & - \\ 
L20-031 & 20:34:48.07 & 60:07:50.2 & 21 & 3.2 & 16.9$\pm$1.2 & 1.2$\pm$0.6 & - & L19-067 & - & - & - \\ 
L20-032 & 20:34:48.23 & 60:06:56.5 & 16 & 5.1 & 3.1$\pm$0.3 & 1.6$\pm$0.4 & - & - & - & - & - \\ 
L20-033 & 20:34:48.43 & 60:08:20.3 & 10 & 2.1 & 4.7$\pm$0.4 & 12.9$\pm$1.1 & C & - & - & - & - \\ 
L20-034 & 20:34:48.43 & 60:10:53.6 & 21 & 4.0 & 29.1$\pm$4.9 & 12.6$\pm$5.4 & C & - & - & - & - \\ 
L20-035 & 20:34:48.62 & 60:09:24.2 & 28 & 1.0 & 17.2$\pm$1.4 & 0.2$\pm$1.6 & - & L19-068 & - & - & - \\ 
L20-036 & 20:34:48.67 & 60:09:34.1 & 18 & 1.2 & 13.2$\pm$0.9 & -0.5$\pm$0.7 & - & - & - & - & - \\ 
L20-037 & 20:34:48.72 & 60:08:23.0 & 19 & 2.0 & 24.9$\pm$2.2 & 1.6$\pm$0.9 & C & L19-069 & - & - & - \\ 
L20-038 & 20:34:48.74 & 60:09:22.3 & 18 & 1.0 & 6.4$\pm$0.8 & 35.2$\pm$6.5 & C & - & - & - & - \\ 
L20-039 & 20:34:49.63 & 60:07:36.6 & 25 & 3.6 & 9.9$\pm$0.7 & -0.4$\pm$0.8 & - & L19-070 & - & - & - \\ 
L20-040 & 20:34:49.73 & 60:12:40.1 & 43 & 7.8 & 10.8$\pm$0.9 & 64.9$\pm$5.0 & C & - & - & - & - \\ 
L20-041 & 20:34:49.98 & 60:09:43.0 & 32 & 1.3 & 10.8$\pm$0.7 & 4.8$\pm$1.4 & - & L19-073 & - & - & - \\ 
L20-042 & 20:34:50.09 & 60:10:23.2 & 20 & 2.7 & 12.1$\pm$0.5 & 1.4$\pm$0.8 & - & - & - & - & - \\ 
L20-043 & 20:34:50.35 & 60:09:45.0 & 15 & 1.3 & 9.8$\pm$1.1 & 1.2$\pm$0.5 & - & L19-074 & - & - & - \\ 
L20-044 & 20:34:50.48 & 60:05:37.7 & 25 & 8.1 & 21.2$\pm$1.3 & -- & - & - & - & - & - \\ 
L20-045 & 20:34:50.78 & 60:07:48.0 & 13 & 3.2 & 67.3$\pm$8.4 & 0.4$\pm$0.7 & - & L19-076 & - & B14-20 & F08-43 \\ 
L20-046 & 20:34:50.93 & 60:10:20.8 & 13 & 2.5 & 217.0$\pm$25.2 & 53.2$\pm$2.8 & C & L19-077 & L97-48 & - & F08-45 \\ 
L20-047 & 20:34:51.01 & 60:10:20.0 & 13 & 2.5 & 15.5$\pm$1.6 & 46.2$\pm$4.4 & C & - & - & - & - \\ 
L20-048 & 20:34:51.21 & 60:09:18.4 & 15 & 0.3 & 11.3$\pm$1.1 & 1.7$\pm$0.6 & - & - & - & - & - \\ 
L20-049 & 20:34:51.25 & 60:09:38.6 & 8 & 1.0 & 6.5$\pm$0.2 & 105.0$\pm$11.2 & C & - & - & - & - \\ 
L20-050 & 20:34:51.29 & 60:09:37.2 & 18 & 0.9 & 9.2$\pm$0.3 & 11.5$\pm$1.8 & C & - & - & - & - \\ 
L20-051 & 20:34:51.38 & 60:09:12.8 & 19 & 0.2 & 22.4$\pm$1.1 & 2.9$\pm$1.3 & - & - & - & - & - \\ 
L20-052 & 20:34:51.40 & 60:07:39.2 & 43 & 3.5 & 97.9$\pm$4.9 & 2.2$\pm$2.5 & - & L19-079 & L97-51 & - & - \\ 
L20-053 & 20:34:51.55 & 60:09:22.7 & 12 & 0.4 & 5.6$\pm$0.4 & 9.0$\pm$2.1 & - & - & - & - & - \\ 
L20-054 & 20:34:51.55 & 60:09:09.1 & 15 & 0.2 & 46.5$\pm$4.1 & 1.9$\pm$1.2 & - & L19-080 & - & - & F08-47 \\ 
L20-055 & 20:34:51.70 & 60:07:30.9 & 15 & 3.9 & 11.8$\pm$1.2 & 3.8$\pm$0.4 & - & - & - & - & - \\ 
L20-056 & 20:34:51.76 & 60:09:11.3 & 19 & 0.1 & 23.2$\pm$1.9 & 10.6$\pm$1.5 & - & - & - & - & - \\ 
L20-057 & 20:34:52.39 & 60:09:11.9 & 9 & 0.1 & 42.2$\pm$3.7 & -18.6$\pm$6.3 & - & - & - & - & - \\ 
L20-058 & 20:34:52.43 & 60:09:17.1 & 7 & 0.1 & 32.9$\pm$5.8 & 23.5$\pm$7.5 & - & - & - & - & - \\ 
L20-059 & 20:34:52.46 & 60:07:27.9 & 45 & 4.0 & 23.1$\pm$1.4 & 3.2$\pm$1.2 & - & L19-082 & - & - & - \\ 
L20-060 & 20:34:52.49 & 60:10:01.7 & 27 & 1.8 & 13.1$\pm$1.0 & -1.4$\pm$1.5 & - & L19-083 & - & - & - \\ 
L20-061 & 20:34:52.56 & 60:10:52.5 & 34 & 3.7 & 10.4$\pm$0.9 & 3.6$\pm$1.2 & - & L19-084 & - & - & - \\ 
L20-062 & 20:34:52.85 & 60:09:30.5 & 12 & 0.6 & 9.4$\pm$1.2 & 169.0$\pm$19.9 & - & - & - & - & - \\ 
L20-063 & 20:34:52.96 & 60:07:53.9 & 15 & 3.0 & 12.5$\pm$2.3 & 12.1$\pm$4.4 & C & - & L97-60: & - & - \\ 
L20-064 & 20:34:53.02 & 60:09:14.3 & 28 & 0.2 & 114.0$\pm$5.7 & 92.5$\pm$9.9 & C & - & - & - & - \\ 
L20-065 & 20:34:53.13 & 60:08:47.7 & 15 & 1.1 & 32.3$\pm$2.1 & 1.7$\pm$1.0 & - & - & L97-63 & - & - \\ 
L20-066 & 20:34:53.18 & 60:10:48.2 & 19 & 3.5 & 5.2$\pm$0.5 & 0.6$\pm$0.7 & - & - & - & - & - \\ 
L20-067 & 20:34:53.38 & 60:10:55.4 & 7 & 3.8 & 10.4$\pm$2.0 & 0.4$\pm$0.3 & - & - & - & - & F08-48 \\ 
L20-068 & 20:34:53.44 & 60:07:15.7 & 12 & 4.5 & 16.7$\pm$1.2 & 0.3$\pm$0.2 & - & - & L97-66 & - & - \\ 
L20-069 & 20:34:53.69 & 60:07:13.8 & 12 & 4.6 & 10.2$\pm$0.7 & 0.9$\pm$1.0 & - & L19-086 & L97-68 & - & - \\ 
L20-070 & 20:34:53.88 & 60:09:11.9 & 24 & 0.5 & 13.5$\pm$1.2 & 6.5$\pm$1.0 & - & - & - & - & F08-50: \\ 
L20-071 & 20:34:53.88 & 60:09:18.7 & 12 & 0.4 & 11.8$\pm$1.9 & 2.9$\pm$1.3 & - & - & L97-69: & - & - \\ 
L20-072 & 20:34:53.93 & 60:10:29.4 & 13 & 2.8 & 6.0$\pm$0.5 & 35.4$\pm$3.7 & C & - & - & - & - \\ 
L20-073 & 20:34:54.27 & 60:11:03.3 & 22 & 4.0 & 13.4$\pm$1.3 & 3.0$\pm$1.0 & - & L19-087 & - & - & - \\ 
L20-074 & 20:34:54.36 & 60:08:53.6 & 11 & 1.0 & 8.7$\pm$0.7 & -2.5$\pm$1.1 & - & - & - & - & - \\ 
L20-075 & 20:34:54.40 & 60:10:55.8 & 25 & 3.8 & 5.9$\pm$0.7 & -1.2$\pm$0.8 & - & L19-088 & - & - & - \\ 
L20-076 & 20:34:54.67 & 60:09:21.9 & 15 & 0.7 & 12.1$\pm$1.1 & 2.6$\pm$0.5 & - & - & - & - & - \\ 
L20-077 & 20:34:54.77 & 60:10:06.6 & 22 & 2.0 & 17.7$\pm$1.2 & 0.3$\pm$0.9 & - & L19-090 & - & - & - \\ 
L20-078 & 20:34:55.03 & 60:09:16.6 & 21 & 0.7 & 13.1$\pm$1.2 & 0.7$\pm$1.1 & - & - & - & - & - \\ 
L20-079 & 20:34:55.13 & 60:10:44.5 & 9 & 3.4 & 5.4$\pm$0.8 & -0.4$\pm$0.4 & - & - & - & - & - \\ 
L20-080 & 20:34:55.29 & 60:09:54.2 & 19 & 1.6 & 7.0$\pm$0.7 & 5.3$\pm$0.9 & - & - & - & - & - \\ 
L20-081 & 20:34:55.52 & 60:07:18.2 & 17 & 4.5 & 9.2$\pm$0.8 & 3.6$\pm$0.7 & - & - & - & - & - \\ 
L20-082 & 20:34:55.90 & 60:07:49.0 & 46 & 3.5 & 17.1$\pm$1.2 & 7.0$\pm$1.4 & - & L19-093 & - & - & - \\ 
L20-083 & 20:34:56.56 & 60:08:19.6 & 17 & 2.5 & 73.5$\pm$10.6 & 8.4$\pm$1.0 & - & L19-094 & - & - & F08-53 \\ 
L20-084 & 20:34:56.81 & 60:08:26.3 & 7 & 2.3 & 29.3$\pm$3.4 & 2.0$\pm$0.7 & C & - & - & - & F08-56: \\ 
L20-085 & 20:34:57.22 & 60:07:46.1 & 46 & 3.7 & 26.3$\pm$1.1 & -0.4$\pm$1.3 & - & - & - & - & - \\ 
L20-086 & 20:34:57.69 & 60:11:02.7 & 26 & 4.1 & 9.5$\pm$0.7 & 0.1$\pm$1.0 & - & - & - & - & - \\ 
L20-087 & 20:34:57.81 & 60:08:09.8 & 12 & 3.0 & 37.1$\pm$3.0 & 2.8$\pm$0.3 & - & L19-095 & - & B14-25 & - \\ 
L20-088 & 20:34:57.91 & 60:07:47.7 & 46 & 3.7 & 25.3$\pm$1.3 & 3.5$\pm$1.2 & - & - & - & - & - \\ 
L20-089 & 20:34:58.28 & 60:08:30.1 & 17 & 2.4 & 19.9$\pm$1.3 & 2.0$\pm$0.7 & - & - & - & - & - \\ 
L20-090 & 20:34:58.75 & 60:09:47.1 & 17 & 1.9 & 11.6$\pm$0.9 & 1.0$\pm$0.4 & - & - & - & - & - \\ 
L20-091 & 20:34:58.81 & 60:10:53.4 & 11 & 3.9 & 14.3$\pm$1.5 & 1.1$\pm$0.7 & - & - & L97-83: & - & - \\ 
L20-092 & 20:34:59.95 & 60:08:35.8 & 6 & 2.6 & 8.9$\pm$1.2 & 1.5$\pm$0.2 & - & - & - & - & - \\ 
L20-093 & 20:35:00.32 & 60:11:45.8 & 36 & 5.8 & 27.0$\pm$1.5 & 8.8$\pm$1.6 & - & L19-097 & - & - & - \\ 
L20-094 & 20:35:00.73 & 60:11:30.7 & 46 & 5.3 & 415.0$\pm$48.2 & 59.1$\pm$4.8 & - & L19-098 & L97-85 & B14-29 & F08-63 \\ 
L20-095 & 20:35:00.80 & 60:11:04.9 & 40 & 4.4 & 19.1$\pm$1.4 & -3.3$\pm$1.4 & - & - & - & - & - \\ 
L20-096 & 20:35:00.89 & 60:08:15.9 & 17 & 3.3 & 7.8$\pm$0.7 & -0.3$\pm$0.6 & - & - & - & - & - \\ 
L20-097 & 20:35:01.15 & 60:12:00.0 & 32 & 6.3 & 6.0$\pm$0.5 & -- & - & L19-099 & - & - & - \\ 
L20-098 & 20:35:02.31 & 60:10:51.3 & 17 & 4.1 & 16.2$\pm$1.9 & 1.2$\pm$0.5 & - & - & - & - & - \\ 
L20-099 & 20:35:03.14 & 60:10:41.7 & 29 & 4.0 & 10.3$\pm$0.8 & 0.8$\pm$1.0 & - & L19-103 & - & - & - \\ 
L20-100 & 20:35:03.31 & 60:10:01.1 & 12 & 3.1 & 8.6$\pm$1.1 & 0.7$\pm$0.8 & - & - & - & - & - \\ 
L20-101 & 20:35:04.04 & 60:11:15.3 & 34 & 5.1 & 9.6$\pm$1.0 & 1.2$\pm$1.2 & - & L19-106 & - & - & - \\ 
L20-102 & 20:35:04.07 & 60:09:54.4 & 18 & 3.2 & 52.0$\pm$2.7 & 24.0$\pm$2.0 & C & - & L97-88 & - & - \\ 
L20-103 & 20:35:04.18 & 60:11:18.2 & 15 & 5.2 & 9.5$\pm$0.8 & 1.2$\pm$0.4 & - & L19-107 & - & - & - \\ 
L20-104 & 20:35:04.21 & 60:09:53.2 & 29 & 3.2 & 15.1$\pm$1.5 & 5.5$\pm$1.9 & - & L19-108 & - & - & - \\ 
L20-105 & 20:35:04.65 & 60:11:17.4 & 7 & 5.2 & 6.8$\pm$0.8 & 0.9$\pm$0.5 & - & - & - & - & F08-67 \\ 
L20-106 & 20:35:04.92 & 60:10:54.1 & 38 & 4.6 & 15.1$\pm$1.0 & 23.6$\pm$2.3 & C & - & - & - & - \\ 
L20-107 & 20:35:05.67 & 60:11:07.5 & 14 & 5.1 & 6.5$\pm$0.6 & -0.1$\pm$0.6 & - & L19-112 & L97-95 & - & - \\ 
L20-108 & 20:35:06.80 & 60:07:58.0 & 47 & 5.0 & 24.8$\pm$1.5 & 5.2$\pm$2.0 & - & L19-113 & - & - & - \\ 
L20-109 & 20:35:06.90 & 60:09:56.7 & 34 & 3.8 & 8.2$\pm$0.7 & 4.0$\pm$1.4 & - & L19-114 & - & - & - \\ 
L20-110 & 20:35:07.25 & 60:09:40.3 & 34 & 3.8 & 8.9$\pm$0.8 & -3.9$\pm$1.2 & - & - & - & - & - \\ 
L20-111 & 20:35:08.12 & 60:11:13.1 & 9 & 5.6 & 115.0$\pm$20.3 & 17.3$\pm$2.3 & C & - & L97-101 & B14-35 & F08-72 \\ 
L20-112 & 20:35:08.22 & 60:11:09.8 & 43 & 5.5 & 17.8$\pm$1.1 & 4.2$\pm$1.9 & - & - & - & - & - \\ 
L20-113 & 20:35:08.38 & 60:07:50.2 & 18 & 5.5 & 13.2$\pm$0.9 & -0.6$\pm$0.7 & - & - & - & - & - \\ 
L20-114 & 20:35:08.90 & 60:07:44.1 & 26 & 5.8 & 8.3$\pm$0.7 & -2.6$\pm$0.9 & - & - & - & - & - \\ 
L20-115 & 20:35:10.63 & 60:10:40.8 & 35 & 5.3 & 8.3$\pm$0.8 & -1.0$\pm$1.2 & - & L19-123 & - & - & - \\ 
L20-116 & 20:35:10.89 & 60:08:56.6 & 15 & 4.9 & 60.9$\pm$9.3 & 6.0$\pm$1.5 & C & L19-124 & - & - & F08-74 \\ 
L20-117 & 20:35:10.94 & 60:11:09.1 & 17 & 5.9 & 17.8$\pm$1.5 & 3.7$\pm$1.1 & - & - & - & - & - \\ 
L20-118 & 20:35:11.02 & 60:08:26.8 & 21 & 5.3 & 10.7$\pm$0.8 & 5.5$\pm$1.7 & - & L19-125 & - & - & - \\ 
L20-119 & 20:35:11.40 & 60:10:32.3 & 28 & 5.3 & 17.1$\pm$1.4 & 41.2$\pm$3.3 & C & - & - & - & - \\ 
L20-120 & 20:35:11.59 & 60:07:41.1 & 29 & 6.4 & 33.6$\pm$2.1 & 6.0$\pm$1.1 & - & L19-127 & - & - & - \\ 
L20-121 & 20:35:11.70 & 60:08:46.6 & 25 & 5.2 & 8.2$\pm$1.3 & 4.9$\pm$1.2 & - & - & - & - & - \\ 
L20-122 & 20:35:11.89 & 60:09:28.4 & 18 & 5.0 & 6.0$\pm$0.7 & 1.1$\pm$0.7 & - & L19-128 & - & - & - \\ 
L20-123 & 20:35:11.99 & 60:10:25.0 & 18 & 5.3 & 17.2$\pm$1.4 & 5.9$\pm$0.6 & - & - & - & - & - \\ 
L20-124 & 20:35:12.21 & 60:09:00.5 & 14 & 5.2 & 5.2$\pm$0.5 & 0.7$\pm$0.4 & - & - & - & - & - \\ 
L20-125 & 20:35:12.59 & 60:09:09.4 & 31 & 5.2 & 28.0$\pm$1.8 & 2.2$\pm$1.5 & - & L19-131 & - & - & - \\ 
L20-126 & 20:35:15.88 & 60:10:08.3 & 31 & 6.1 & 11.1$\pm$0.5 & -0.4$\pm$1.2 & - & - & - & - & - \\ 
L20-127 & 20:35:16.34 & 60:11:02.7 & 18 & 6.8 & 6.1$\pm$1.0 & -- & - & - & - & - & - \\ 
L20-128 & 20:35:16.93 & 60:11:05.1 & 33 & 7.0 & 20.0$\pm$1.2 & -- & - & L19-135 & - & - & - \\ 
L20-129 & 20:35:20.03 & 60:09:33.7 & 16 & 7.0 & 34.0$\pm$3.7 & 2.7$\pm$0.6 & - & L19-138 & - & - & F08-82 \\ 
L20-130 & 20:35:20.78 & 60:09:52.4 & 29 & 7.2 & 9.4$\pm$0.7 & -0.5$\pm$1.3 & - & L19-139 & - & - & - \\ 
L20-131 & 20:35:21.14 & 60:09:51.3 & 18 & 7.3 & 5.5$\pm$0.3 & 2.2$\pm$0.5 & C & - & - & - & F08-83 \\ 
L20-132 & 20:35:24.72 & 60:10:00.0 & 8 & 8.2 & 9.8$\pm$1.0 & 5.1$\pm$2.6 & C & - & - & - & - \\ 
\enddata
\tablenotetext{a}{Flux in units of 10$^{-17}$ ergs cm$^{-2}$ s$^{-1}$.  {A negative (Pa$\beta$)  flux implies that the average count rate in the source region was less than that in the chosen background region.} }
\tablenotetext{b}{If labeled C, the [Fe II] sources is in a relatively complex region with extended H$\alpha$ and/or Pa$\beta$ See text.}
\tablenotetext{c}{There are two catalog sources, L20-20 and L20-22 associated with the radio source L97-26. }
\tablenotetext{d}{A trailing colon indicates that while a radio or X-ray source was within 2 \arcsec\ of the [Fe II] source. While close, the positional match was not as precise as most of the rest of the sample.}
\tablenotetext{}{References: Optical: \citet[][L19]{long19}; radio: \citet{lacey01}; \citet{bruursema14} is previous ground-based \feii\ survey; X-ray: \citet[][F08]{fridriksson08}.}
\label{ngc6946_tab1}
\end{deluxetable}
 \end{longrotatetable}
 
 \startlongtable
\begin{deluxetable}{rccrrcccc}
\tablecaption{Optical SNRs}
\tablehead{
\colhead{Name} &
\colhead{R.A.} &
\colhead{Decl.} &
\colhead{D} &
\colhead{R} &
\colhead{Confirmed$^a$} &
\colhead{[Fe II] cat.$^b$} &
\colhead{[Fe II] flux$^c$} &
\colhead{Pa$\beta$ flux$^c$}
\\
\colhead{} &
\colhead{(J2000)} &
\colhead{(J2000)} &
\colhead{(pc)} &
\colhead{(kpc)} &
\colhead{} &
\colhead{} &
\colhead{} &
\colhead{}
}
\startdata
L19-001 & 20:34:15.00 & 60:10:44.3 & 71 & 10.4 & no & - & -- & -- \\ 
L19-002 & 20:34:15.47 & 60:07:31.6 & 88 & 9.6 & - & - & -- & -- \\ 
L19-003 & 20:34:15.76 & 60:08:26.0 & 337 & 9.2 & - & - & -- & -- \\ 
L19-004 & 20:34:16.41 & 60:08:27.3 & 62 & 9.0 & no & - & -- & -- \\ 
L19-005 & 20:34:16.68 & 60:07:30.8 & 109 & 9.3 & no & - & -- & -- \\ 
L19-006 & 20:34:17.56 & 60:10:58.2 & 213 & 10.1 & yes & - & -- & -- \\ 
L19-007 & 20:34:17.96 & 60:10:00.6 & 99 & 9.1 & yes & - & -- & -- \\ 
L19-008 & 20:34:18.42 & 60:10:47.0 & 298 & 9.7 & - & - & -- & -- \\ 
L19-009 & 20:34:18.84 & 60:11:08.9 & 71 & 10.0 & yes & - & -- & -- \\ 
L19-010 & 20:34:19.17 & 60:08:57.5 & 118 & 8.3 & yes & - & -- & -- \\ 
L19-011 & 20:34:20.58 & 60:09:06.7 & 32 & 8.0 & yes & - & -- & -- \\ 
L19-012 & 20:34:21.95 & 60:08:58.0 & 36 & 7.6 & - & - & -- & 1.4$\pm$1.1 \\ 
L19-013 & 20:34:22.69 & 60:06:13.4 & 53 & 9.4 & yes & - & -- & -- \\ 
L19-014 & 20:34:23.38 & 60:08:18.4 & 42 & 7.3 & yes & L20-001 & 26.1$\pm$1.4 & 4.3$\pm$1.4 \\ 
L19-015 & 20:34:23.39 & 60:11:35.3 & 106 & 9.6 & yes & - & -- & -- \\ 
L19-016 & 20:34:24.44 & 60:11:25.9 & 47 & 9.1 & yes & - & -- & -- \\ 
L19-017 & 20:34:24.91 & 60:09:46.4 & 129 & 7.2 & no & - & -- & -11.1$\pm$13.7 \\ 
L19-018 & 20:34:25.39 & 60:08:55.9 & 23 & 6.7 & - & no & -1.9$\pm$0.8 & 2.0$\pm$1.2 \\ 
L19-019 & 20:34:26.01 & 60:11:10.7 & 74 & 8.4 & yes & - & -- & -- \\ 
L19-020 & 20:34:26.06 & 60:13:22.8 & 92 & 12.2 & - & - & -- & -- \\ 
L19-021 & 20:34:26.18 & 60:10:11.9 & 136 & 7.2 & - & - & -- & -- \\ 
L19-022 & 20:34:27.67 & 60:11:12.4 & 47 & 8.1 & - & - & -- & -- \\ 
L19-023 & 20:34:28.22 & 60:11:37.7 & 45 & 8.7 & - & - & -- & -- \\ 
L19-024 & 20:34:28.32 & 60:13:22.0 & 126 & 11.8 & - & - & -- & -- \\ 
L19-025 & 20:34:28.32 & 60:07:04.2 & 54 & 7.2 & yes & - & -- & -- \\ 
L19-026 & 20:34:28.36 & 60:08:09.1 & 43 & 6.2 & - & no & 3.3$\pm$0.8 & -3.4$\pm$1.6 \\ 
L19-027 & 20:34:28.42 & 60:07:33.5 & 82 & 6.7 & - & no & 1.5$\pm$1.8 & -- \\ 
L19-028 & 20:34:28.88 & 60:07:45.2 & 33 & 6.4 & no & L20-002 & 4.9$\pm$1.0 & 21.7$\pm$2.2 \\ 
L19-029 & 20:34:29.17 & 60:10:51.1 & 79 & 7.3 & - & - & -- & -- \\ 
L19-030 & 20:34:30.11 & 60:10:24.3 & 48 & 6.5 & yes & - & -- & -1.1$\pm$1.6 \\ 
L19-031 & 20:34:31.64 & 60:10:27.8 & 35 & 6.2 & yes & L20-004 & 10.4$\pm$0.9 & -0.3$\pm$1.2 \\ 
L19-032 & 20:34:32.59 & 60:10:27.7 & 23 & 6.0 & no & L20-006 & 18.0$\pm$1.6 & -0.8$\pm$0.8 \\ 
L19-033 & 20:34:33.04 & 60:11:25.4 & 54 & 7.4 & yes & - & -- & 0.4$\pm$3.3 \\ 
L19-034 & 20:34:33.30 & 60:09:46.5 & 31 & 5.1 & - & L20-008 & 19.3$\pm$1.4 & 0.5$\pm$0.8 \\ 
L19-035 & 20:34:33.62 & 60:09:51.9 & 31 & 5.1 & - & L20-009 & 15.3$\pm$1.1 & 1.2$\pm$1.1 \\ 
L19-036 & 20:34:33.83 & 60:09:25.1 & 88 & 4.7 & yes & no & 15.1$\pm$3.5 & 15.8$\pm$4.5 \\ 
L19-037 & 20:34:36.63 & 60:11:34.4 & 86 & 7.0 & yes & - & -- & 0.5$\pm$3.1 \\ 
L19-038 & 20:34:37.37 & 60:07:15.0 & 82 & 5.4 & yes & no & 6.3$\pm$2.6 & 6.3$\pm$6.7 \\ 
L19-039 & 20:34:37.43 & 60:11:31.3 & 44 & 6.8 & yes & L20-012 & 9.6$\pm$0.7 & 0.6$\pm$1.1 \\ 
L19-040 & 20:34:37.75 & 60:08:52.5 & 31 & 3.6 & yes & no & 2.7$\pm$1.5 & -5.6$\pm$2.9 \\ 
L19-041 & 20:34:37.79 & 60:11:54.3 & 20 & 7.4 & yes & - & -- & 0.5$\pm$0.5 \\ 
L19-042 & 20:34:37.96 & 60:07:22.0 & 33 & 5.1 & yes & L20-015 & 15.8$\pm$1.0 & 2.2$\pm$1.0 \\ 
L19-043 & 20:34:38.36 & 60:06:09.4 & 158 & 7.3 & - & - & -- & -- \\ 
L19-044 & 20:34:38.90 & 60:06:57.7 & 112 & 5.7 & yes & no & 4.3$\pm$4.1 & 1.5$\pm$7.8 \\ 
L19-045 & 20:34:39.11 & 60:09:18.5 & 23 & 3.3 & - & L20-016 & 11.1$\pm$0.6 & 0.8$\pm$0.8 \\ 
L19-046 & 20:34:39.16 & 60:08:13.7 & 15 & 3.7 & yes & L20-018 & 18.5$\pm$1.6 & 0.7$\pm$0.8 \\ 
L19-047 & 20:34:39.65 & 60:07:26.0 & 34 & 4.8 & - & no & 1.0$\pm$0.6 & 0.7$\pm$1.1 \\ 
L19-048 & 20:34:40.66 & 60:06:53.2 & 50 & 5.7 & yes & L20-019 & 9.8$\pm$1.4 & 0.1$\pm$1.5 \\ 
L19-049 & 20:34:40.72 & 60:08:33.2 & 57 & 3.1 & yes & no & 7.5$\pm$1.1 & -3.1$\pm$2.3 \\ 
L19-050 & 20:34:41.01 & 60:05:57.5 & 40 & 7.5 & - & - & -- & -- \\ 
L19-051 & 20:34:41.31 & 60:11:12.6 & 43 & 5.5 & yes & L20-021 & 7.9$\pm$0.9 & 3.3$\pm$0.8 \\ 
L19-052 & 20:34:41.32 & 60:04:54.9 & 115 & 9.7 & - & - & -- & -- \\ 
L19-053 & 20:34:41.50 & 60:11:29.8 & 23 & 6.1 & yes & no & 1.5$\pm$0.5 & 3.1$\pm$0.9 \\ 
L19-054 & 20:34:41.89 & 60:05:50.0 & 84 & 7.8 & yes & - & -- & -- \\ 
L19-055 & 20:34:42.42 & 60:09:15.9 & 31 & 2.5 & yes & no & 3.5$\pm$0.5 & 0.3$\pm$0.8 \\ 
L19-056 & 20:34:43.08 & 60:11:39.3 & 58 & 6.2 & no & no & 5.7$\pm$1.2 & 10.5$\pm$1.9 \\ 
L19-057 & 20:34:43.30 & 60:10:11.1 & 75 & 3.3 & - & no & 9.0$\pm$1.9 & 4.6$\pm$3.3 \\ 
L19-058 & 20:34:43.50 & 60:07:51.6 & 48 & 3.5 & - & no & 5.7$\pm$1.0 & 2.2$\pm$2.0 \\ 
L19-059 & 20:34:43.97 & 60:08:24.0 & 32 & 2.6 & yes & L20-025 & 8.9$\pm$0.8 & 0.4$\pm$0.9 \\ 
L19-060 & 20:34:44.59 & 60:08:17.0 & 14 & 2.7 & yes & L20-026 & 8.6$\pm$0.5 & 0.2$\pm$0.5 \\ 
L19-061 & 20:34:45.13 & 60:12:36.4 & 76 & 8.0 & yes & no & 10.6$\pm$1.6 & 0.6$\pm$3.3 \\ 
L19-062 & 20:34:45.64 & 60:07:20.8 & 42 & 4.3 & yes & no & 5.8$\pm$1.1 & -1.5$\pm$2.0 \\ 
L19-063 & 20:34:46.93 & 60:12:19.7 & 29 & 7.2 & yes & no & 1.4$\pm$0.4 & 1.2$\pm$0.8 \\ 
L19-064 & 20:34:47.16 & 60:08:20.0 & 26 & 2.2 & yes & no & 1.4$\pm$0.7 & -0.6$\pm$1.1 \\ 
L19-065 & 20:34:47.37 & 60:08:22.3 & 29 & 2.1 & yes & L20-028 & 25.7$\pm$3.3 & 2.8$\pm$0.9 \\ 
L19-066 & 20:34:47.75 & 60:09:58.7 & 70 & 2.1 & yes & no & 12.1$\pm$1.9 & -3.6$\pm$3.9 \\ 
L19-067 & 20:34:48.07 & 60:07:50.2 & 21 & 3.2 & yes & L20-031 & 16.9$\pm$1.2 & 1.2$\pm$0.6 \\ 
L19-068 & 20:34:48.62 & 60:09:24.2 & 28 & 1.0 & yes & L20-035 & 17.2$\pm$1.4 & 0.2$\pm$1.6 \\ 
L19-069 & 20:34:48.72 & 60:08:23.0 & 20 & 2.0 & yes & L20-037 & 24.9$\pm$2.2 & 1.6$\pm$0.9 \\ 
L19-070 & 20:34:49.63 & 60:07:36.6 & 26 & 3.6 & yes & L20-039 & 9.9$\pm$0.7 & -0.4$\pm$0.8 \\ 
L19-071 & 20:34:49.76 & 60:09:41.1 & 32 & 1.2 & - & no & 2.8$\pm$1.0 & -1.1$\pm$0.9 \\ 
L19-072 & 20:34:49.91 & 60:07:52.9 & 16 & 3.0 & yes & no & 0.8$\pm$0.4 & -0.3$\pm$0.5 \\ 
L19-073 & 20:34:49.98 & 60:09:43.0 & 33 & 1.3 & - & L20-041 & 10.8$\pm$0.7 & 4.8$\pm$1.4 \\ 
L19-074 & 20:34:50.35 & 60:09:45.0 & 16 & 1.3 & yes & L20-043 & 9.8$\pm$1.1 & 1.2$\pm$0.5 \\ 
L19-075 & 20:34:50.36 & 60:09:51.6 & 26 & 1.5 & - & no & 6.7$\pm$0.8 & -1.2$\pm$0.8 \\ 
L19-076 & 20:34:50.78 & 60:07:48.0 & 14 & 3.2 & yes & L20-045 & 67.3$\pm$8.4 & 0.4$\pm$0.7 \\ 
L19-077 & 20:34:50.93 & 60:10:20.8 & 14 & 2.6 & - & L20-046 & 217.0$\pm$25.2 & 53.2$\pm$2.8 \\ 
L19-078 & 20:34:51.29 & 60:05:20.5 & 204 & 8.7 & - & - & -- & -- \\ 
L19-079 & 20:34:51.44 & 60:07:39.1 & 27 & 3.5 & yes & L20-052 & 97.9$\pm$4.9 & 2.2$\pm$2.5 \\ 
L19-080 & 20:34:51.55 & 60:09:09.1 & 16 & 0.2 & yes & L20-054 & 46.5$\pm$4.1 & 1.9$\pm$1.2 \\ 
L19-081 & 20:34:51.64 & 60:09:56.7 & 45 & 1.6 & no & no & -5.3$\pm$6.2 & -9.3$\pm$14.0 \\ 
L19-082 & 20:34:52.45 & 60:07:28.0 & 47 & 4.0 & yes & L20-059 & 23.1$\pm$1.4 & 3.2$\pm$1.2 \\ 
L19-083 & 20:34:52.49 & 60:10:01.7 & 27 & 1.8 & yes & L20-060 & 13.1$\pm$1.0 & -1.4$\pm$1.5 \\ 
L19-084 & 20:34:52.55 & 60:10:52.4 & 36 & 3.7 & yes & L20-061 & 10.4$\pm$0.9 & 3.6$\pm$1.2 \\ 
L19-085 & 20:34:53.10 & 60:08:13.9 & 49 & 2.3 & yes & no & 8.7$\pm$1.1 & 3.9$\pm$1.8 \\ 
L19-086 & 20:34:53.69 & 60:07:13.8 & 12 & 4.6 & yes & L20-069 & 10.2$\pm$0.7 & 0.9$\pm$1.0 \\ 
L19-087 & 20:34:54.27 & 60:11:03.3 & 23 & 4.0 & yes & L20-073 & 13.4$\pm$1.3 & 3.0$\pm$1.0 \\ 
L19-088 & 20:34:54.40 & 60:10:55.8 & 25 & 3.8 & yes & L20-075 & 5.9$\pm$0.7 & -1.2$\pm$0.8 \\ 
L19-089 & 20:34:54.54 & 60:05:08.7 & 269 & 9.3 & yes & - & -- & -- \\ 
L19-090 & 20:34:54.77 & 60:10:06.6 & 23 & 2.0 & yes & L20-077 & 17.7$\pm$1.2 & 0.3$\pm$0.9 \\ 
L19-091 & 20:34:54.90 & 60:10:34.5 & 51 & 3.0 & yes & no & 6.2$\pm$1.9 & -4.0$\pm$2.0 \\ 
L19-092 & 20:34:55.64 & 60:11:13.6 & 32 & 4.4 & - & no & 1.5$\pm$1.2 & 0.8$\pm$1.7 \\ 
L19-093 & 20:34:55.90 & 60:07:49.0 & 46 & 3.5 & yes & L20-082 & 17.1$\pm$1.2 & 7.0$\pm$1.4 \\ 
L19-094 & 20:34:56.56 & 60:08:19.6 & 17 & 2.5 & yes & L20-083 & 73.5$\pm$10.6 & 8.4$\pm$1.0 \\ 
L19-095 & 20:34:57.81 & 60:08:09.8 & 13 & 3.0 & yes & L20-087 & 37.1$\pm$3.0 & 2.8$\pm$0.3 \\ 
L19-096 & 20:34:58.47 & 60:08:01.5 & 27 & 3.3 & yes & no & 5.9$\pm$0.6 & 1.2$\pm$0.7 \\ 
L19-097 & 20:35:00.32 & 60:11:45.8 & 36 & 5.8 & yes & L20-093 & 27.0$\pm$1.5 & 8.8$\pm$1.6 \\ 
L19-098 & 20:35:00.73 & 60:11:30.7 & 46 & 5.3 & yes & L20-094 & 415.0$\pm$48.2 & 59.1$\pm$4.8 \\ 
L19-099 & 20:35:01.15 & 60:12:00.0 & 33 & 6.3 & yes & L20-097 & 6.0$\pm$0.5 & -- \\ 
L19-100 & 20:35:02.24 & 60:11:05.2 & 70 & 4.6 & yes & no & 1.2$\pm$1.9 & -6.5$\pm$2.8 \\ 
L19-101 & 20:35:02.37 & 60:06:31.4 & 99 & 7.0 & yes & no & 9.7$\pm$2.5 & 3.6$\pm$4.4 \\ 
L19-102 & 20:35:02.95 & 60:11:27.2 & 65 & 5.3 & yes & no & 8.6$\pm$1.4 & 10.0$\pm$2.6 \\ 
L19-103 & 20:35:03.14 & 60:10:41.7 & 29 & 4.0 & yes & L20-099 & 10.3$\pm$0.8 & 0.8$\pm$1.0 \\ 
L19-104 & 20:35:03.22 & 60:05:28.0 & 262 & 9.3 & yes & - & -- & -- \\ 
L19-105 & 20:35:03.59 & 60:06:23.3 & 114 & 7.4 & - & no & 5.3$\pm$3.3 & -4.2$\pm$6.2 \\ 
L19-106 & 20:35:04.04 & 60:11:15.3 & 34 & 5.1 & yes & L20-101 & 9.6$\pm$1.0 & 1.2$\pm$1.2 \\ 
L19-107 & 20:35:04.18 & 60:11:18.2 & 15 & 5.2 & - & L20-103 & 9.5$\pm$0.8 & 1.2$\pm$0.4 \\ 
L19-108 & 20:35:04.21 & 60:09:53.2 & 29 & 3.2 & no & L20-104 & 15.1$\pm$1.5 & 5.5$\pm$1.9 \\ 
L19-109 & 20:35:04.25 & 60:06:51.9 & 23 & 6.5 & yes & no & 1.8$\pm$0.5 & -0.4$\pm$0.6 \\ 
L19-110 & 20:35:04.98 & 60:05:32.9 & 88 & 9.3 & - & - & -- & -- \\ 
L19-111 & 20:35:05.68 & 60:10:00.7 & 68 & 3.6 & yes & no & 4.9$\pm$2.0 & 11.5$\pm$3.4 \\ 
L19-112 & 20:35:05.67 & 60:11:07.5 & 14 & 5.1 & yes & L20-107 & 6.5$\pm$0.6 & -0.1$\pm$0.6 \\ 
L19-113 & 20:35:06.81 & 60:07:58.1 & 46 & 5.0 & yes & L20-108 & 24.8$\pm$1.5 & 5.2$\pm$2.0 \\ 
L19-114 & 20:35:06.93 & 60:09:56.7 & 33 & 3.9 & yes & L20-109 & 8.2$\pm$0.7 & 4.0$\pm$1.4 \\ 
L19-115 & 20:35:07.05 & 60:05:57.3 & 117 & 8.8 & - & - & -- & -- \\ 
L19-116 & 20:35:08.80 & 60:06:03.0 & 144 & 8.8 & - & - & -- & -- \\ 
L19-117 & 20:35:08.89 & 60:10:13.0 & 74 & 4.5 & - & no & -1.5$\pm$1.9 & 0.2$\pm$3.3 \\ 
L19-118 & 20:35:09.48 & 60:09:12.8 & 96 & 4.4 & no & no & -3.0$\pm$3.5 & 25.2$\pm$7.1 \\ 
L19-119 & 20:35:09.60 & 60:12:29.8 & 39 & 8.0 & - & - & -- & -- \\ 
L19-120 & 20:35:09.87 & 60:06:13.3 & 92 & 8.6 & - & - & -- & -- \\ 
L19-121 & 20:35:10.21 & 60:06:26.7 & 92 & 8.3 & yes & - & -- & -- \\ 
L19-122 & 20:35:10.55 & 60:06:41.3 & 98 & 7.9 & - & - & -- & -- \\ 
L19-123 & 20:35:10.63 & 60:10:41.0 & 34 & 5.3 & yes & L20-115 & 8.3$\pm$0.8 & -1.0$\pm$1.2 \\ 
L19-124 & 20:35:10.89 & 60:08:56.6 & 15 & 4.9 & no & L20-116 & 60.9$\pm$9.3 & 6.0$\pm$1.5 \\ 
L19-125 & 20:35:11.02 & 60:08:26.8 & 22 & 5.3 & yes & L20-118 & 10.7$\pm$0.8 & 5.5$\pm$1.7 \\ 
L19-126 & 20:35:11.42 & 60:11:11.9 & 84 & 6.1 & yes & no & 3.3$\pm$3.0 & -- \\ 
L19-127 & 20:35:11.59 & 60:07:41.1 & 29 & 6.4 & yes & L20-120 & 33.6$\pm$2.1 & 6.0$\pm$1.1 \\ 
L19-128 & 20:35:11.90 & 60:09:28.4 & 19 & 5.0 & yes & L20-122 & 6.0$\pm$0.7 & 1.1$\pm$0.7 \\ 
L19-129 & 20:35:11.94 & 60:04:03.8 & 175 & 13.3 & - & - & -- & -- \\ 
L19-130 & 20:35:12.25 & 60:06:37.6 & 160 & 8.3 & - & - & -- & -- \\ 
L19-131 & 20:35:12.59 & 60:09:09.4 & 31 & 5.2 & yes & L20-125 & 28.0$\pm$1.8 & 2.2$\pm$1.5 \\ 
L19-132 & 20:35:13.61 & 60:08:58.9 & 153 & 5.5 & yes & no & 8.3$\pm$5.3 & -6.3$\pm$10.9\\ 
L19-133 & 20:35:14.45 & 60:07:12.7 & 64 & 7.7 & yes & - & -- & -- \\ 
L19-134 & 20:35:16.52 & 60:07:50.1 & 65 & 7.3 & - & - & -- & -- \\ 
L19-135 & 20:35:16.93 & 60:11:05.1 & 33 & 7.0 & yes & L20-128 & 20.0$\pm$1.2 & -- \\ 
L19-136 & 20:35:17.39 & 60:10:28.4 & 62 & 6.6 & yes & no & 1.5$\pm$1.4 & 9.2$\pm$2.8 \\ 
L19-137 & 20:35:17.47 & 60:07:20.4 & 144 & 8.2 & - & - & -- & -- \\ 
L19-138 & 20:35:20.03 & 60:09:33.7 & 17 & 7.0 & yes & L20-129 & 34.0$\pm$3.7 & 2.7$\pm$0.6 \\ 
L19-139 & 20:35:20.78 & 60:09:52.4 & 29 & 7.2 & - & L20-130 & 9.4$\pm$0.7 & -0.5$\pm$1.3 \\ 
L19-140 & 20:35:21.12 & 60:08:44.0 & 107 & 7.6 & yes & - & -- & -- \\ 
L19-141 & 20:35:23.02 & 60:08:21.2 & 76 & 8.3 & yes & - & -- & -- \\ 
L19-142 & 20:35:23.66 & 60:08:47.7 & 131 & 8.2 & no & - & -- & -- \\ 
L19-143 & 20:35:24.23 & 60:07:42.4 & 133 & 9.2 & no & - & -- & -- \\ 
L19-144 & 20:35:24.61 & 60:06:57.1 & 110 & 10.3 & - & - & -- & -- \\ 
L19-145 & 20:35:25.24 & 60:07:26.9 & 141 & 9.8 & - & - & -- & -- \\ 
L19-146 & 20:35:25.53 & 60:07:51.3 & 99 & 9.4 & - & - & -- & -- \\ 
L19-147 & 20:35:26.11 & 60:08:43.0 & 104 & 8.8 & yes & - & -- & -- \\ 
\enddata
\tablenotetext{a}{``Yes" implies spectroscopically confirmed to have \sii:H$\alpha$ $>$ 0.4; ``no" means a spectroscopic observation was made, but the ratio was lower than 0.4.}
\tablenotetext{b}{If in the [Fe II] catalog, the name of the object is given. Entries labeled ``no" indicate objects that are not in the [Fe II] catalog but that {\em are} in the region observed in [Fe II].}
\tablenotetext{c}{Flux in units of 10$^{-17}$ ergs cm$^{-2}$ s$^{-1}$. {A negative  flux implies that the average count rate in the source region was less than that in the chosen background region.} Values for [Fe II] catalog objects are the same as in Table 1.}
\label{ngc6946_tab2}
\end{deluxetable}

\clearpage
 \begin{deluxetable}{ccccccc}
\tablewidth{0pt}
 \tablecaption{Coincidences between SNR Candidates from Different Surveys}
 \label{tab:overlap}
 \tablehead{
\colhead{} & \colhead {\phn \feii$^a$ \vspace{-0.4cm}} & & & \colhead{\phn X-ray$^{b,c}$} & \colhead{Radio$^{b,d}$} & \colhead{Total SNR$^e$} \\
\colhead{\vspace{-0.4cm}} & & \colhead{L19$^b$} & \colhead{MF97$^b$} & & & \\
\colhead{} \vspace{0.05cm}& \colhead{Sources} &  \colhead {} & \colhead {} & \colhead {Sources} & \colhead{SNRs} & \colhead{Candidates}
}
 \startdata 
 \feii\ Sources$^a$ & 132  & 54 & 15 & 15 & 14 & 132\\
 L19$^b$  & & 92 &19  &   \phn7      &      \phn7  & 147\\
MF97$^b$  & &  & 19 & \phn 2 &\phn2 & \phn27\\
 X-ray Sources$^{b,c}$  &  & & & 67 & \phn6 & \phn--- \\
Radio SNRs$^{b,d}$  & & & & & 30 &\phn 35\\
Total SNR Candidates$^e$ & & & & &  &225
  \enddata
\tablenotetext{a}{\feii\ sources from this survey (Table 1).}
\tablenotetext{b}{Only the sources that fall within the footprint of the WFC3 \feii\ survey.}
\tablenotetext{c}{X-ray sources from the {\em Chandra} survey by \citet{fridriksson08}, which includes a total of 90 objects.  Their catalog includes all X-ray sources, not only suggested SNRs, hence there is no entry in the right-hand column.}
\tablenotetext{d}{Radio SNRs from  \citet{lacey01}, who list a total of 35 SNR candidates.}
\tablenotetext{e}{Includes confirmed or suggested SNR \feii\ candidates (Table 1) plus optical ones from L19 throughout \gal, whether or not they are within the \feii\ survey footprint. It does {\em not} include  21 additional radio sources, identified by \citet{lacey01} as possible SNRs, that have no optical or IR confirmation. }
\end{deluxetable}


%


\end{document}